\documentclass[aps,pre,showpacs,twocolumn,a4paper,floatfix,superscriptaddress]{revtex4-1}
\usepackage{graphicx}% Include figure files
\usepackage{dcolumn}% Align table columns on decimal point
\usepackage{bm}% bold math
\usepackage{color}% bold math
\usepackage{amssymb}
\usepackage{amsmath}
\usepackage{natbib}
\usepackage{multirow,setspace,times,amssymb,amsmath,graphicx,color,rotating,subfigure,url}

\begin{document}

\title{Events Determine Spreading Patterns: Information Transmission via Internal and External Influences on Social Networks}

\author{Chuang Liu}
\affiliation{Alibaba Research Center for Complexity Sciences, Hangzhou Normal University, Hangzhou 311121, P. R. China}
\author{Xiu-Xiu Zhan}
\affiliation{Alibaba Research Center for Complexity Sciences, Hangzhou Normal University, Hangzhou 311121, P. R. China}
\affiliation{Department of Mathematics, North University of China, Taiyuan 030051, P. R. China}
\author{Zi-Ke Zhang}
\thanks{zhangzike@gmail.com}
\affiliation{Alibaba Research Center for Complexity Sciences, Hangzhou Normal University, Hangzhou 311121, P. R. China}
\affiliation{Alibaba Research Institute, Hangzhou, 311121, P. R. China}
\author{Gui-Quan Sun}
\affiliation{Department of Mathematics, North University of China, Taiyuan 030051, P. R. China}
\affiliation{Complex Systems Research Center, Shanxi University, Taiyuan, 030006, P. R. China}
\author{Pak Ming Hui}
\affiliation{Department of Physics, The Chinese University of Hong Kong, Shatin, New Territories, Hong Kong, China}

\begin{abstract}
Recently, information transmission models motivated by the classical epidemic propagation, have been applied to a wide-range of social systems, generally assume that information mainly transmits among individuals via peer-to-peer interactions on social networks. In this paper, we consider one more approach for users to get information: the out-of-social-network influence. Empirical analyses of eight typical events' diffusion on a very large micro-blogging system, \emph{Sina Weibo}, show that the external influence has significant impact on information spreading along with social activities. In addition, we propose a theoretical model to interpret the spreading process via both internal and external channels, considering three essential properties: (i) memory effect; (ii) role of spreaders; and (iii) non-redundancy of contacts. Experimental and mathematical results indicate that the information indeed spreads much quicker and broader with mutual effects of the internal and external influences. More importantly, the present model reveals that the event characteristic would highly determine the essential spreading patterns once the network structure is established. The results may shed some light on the in-depth understanding of the underlying dynamics of information transmission on real social networks.
\end{abstract}


\maketitle

\section{Introduction}
%\label{Sec:Introuction}

How social networks affect information transmission or information spreading is a pressing problem.  Among the spreading phenomena studied in recent years are
news~\cite{Chen-Chen-Gunnell-Yip-2013-PlosOne} and rumors spreading~\cite{Zhang-Zhou-Zhang-Guan-Zhou-2013-PRE,Moreno-Nekovee-Pacheco-2004-PRE}, innovation
diffusion~\cite{Montanari-Saberi-2010-PNAS,Peres2014PA}, human behaviors~\cite{Centola-2010-Science,Blansky-Kavanaugh-Boothroyd-Benson-Gallagher-Endress-Sayama-2013-PLOSONE},
and culture transmission~\cite{Allen-Weinrich-Hoppitt-Rendell-2013-Science,Dybiec-Mitarai-Sneppen-2012-PRE}.
The structure of a network is crucial in determining the spreading pattern and thus widely
studied~\cite{Nematzadeh-Ferrara-Flammini-Ahn-2014-PRL,Nagata-Shirayama-2012-PA},
with the critical phenomenon on network topology \cite{Pastor-Satorras-Vespignani-2001-PRL,Castellano-Pastor-Satorras-2010-PRL},
identification of influential spreaders \cite{Kitsak-Gallos-Havlin-Liljeros-Muchnik-Stanley-Makse-2010-NP,Aral-2011-MS,Chen2013PO},
and spreading dynamics on adaptive networks \cite{Gross-Dlima-Blasius-2006-PRL,Zhou2014PA} being the focuses.
With the increasing availability of real and good-quality data for analysis, the propagation paths
~\cite{Shen-Wang-Fan-Di-Lai-2014-NC,Gomez-Rodriguez-Leskovec-Scholkopf-2013-WSDM},
patterns of human activities ~\cite{Iribarren-Moro-2009-PRL,Karsai-Kivela-Pan-Kaski-Kertesz-Barabasi-Saramaki-2011-PRE}
and locating the source ~\cite{Pinto-Thiran-Vetterli-2012-PRL,Comin-Costa-2011-PRE} also
become the hot spots in studying spreading dynamics.

Theoretical studies on information spreading are mostly carried out within the framework of epidemic spreading~\cite{Pastor-Satorras-Vespignani-2001-PRL}, where the propagation is regarded as a sequence of social interactions between infected and susceptible individuals~\cite{Lloyd-May-2001-Science,Newman-2002-PRE}.
Simulation results from such models, however, are very different from those observed in empirical analyses on real data~\cite{Goel-Watts-Goldstein-2012-EC} as information spreading carries its special features.  Normally, an online individual is unlikely to forward the same piece of news to his friends repeatedly, but s/he could infect (be infected by) a friend the same disease more than once~\cite{Lu-Chen-Zhou-2011-NJP}.  The memory~\cite{Dodds-Watts-2004-PRL} and temporal effects \cite{HuangJM2014SR} are also significantly different, with previous behaviors having grave implications for the information spreading process.  In addition, the information content \cite{Crane-Sornette-2008-PNAS} and timeliness \cite{Wu-Huberman-2007-PNAS} would generate spreading patterns that are very different from epidemic propagation.

\begin{table*}[!ht]
 \centering
 \caption{\label{TB:Data} Basic statistics of the eight representative events.  $day$ represents the date when the corresponding event happens, $N_m$ represents the number of new tweets talking about the corresponding event, $N_r$ represents the total number of new tweets and retweets about the event, and $\langle N_r\rangle$ represents the average retweet number of each tweet. }
  \newcommand{\minitab}[2][1]{\begin{tabular}{#1}#2\end{tabular}}
 \begin{tabular}{ccccccccc}
 \hline
   No. & Events & $day$ & $N_m$ & $N_r$ & $\langle N_r\rangle$\\\hline
   a& Wenzhou Train Collision & 23/Jul/2011 & 91,876 & 448,536 & 4.88 \\
   %b& Death of Osama Bin Laden & 01/May/2011 & 167,597 & 588,698 & 3.51 \\
   b& Yao Ming Retire & 20/Jul/2011 & 39,707 & 109,159 & 2.74 \\
   \multirow{2}*{c} & \multirow{2}*{\minitab[c]{Case of Running Fast Car \\in Hebei University}} & \multirow{2}*{16/Oct/2010} & \multirow{2}*{107,674} & \multirow{2}*{488,991} & \multirow{2}*{4.54} \\\\
   \multirow{2}*{d} & \multirow{2}*{\minitab[c]{Tang Jun Education \\Qualification Fake}} & \multirow{2}*{01/Jul/2010} & \multirow{2}*{122,088} & \multirow{2}*{408,301} & \multirow{2}*{3.34} \\\\
   e & Yushu Earthquake & 14/Apr/2010 & 47,441 & 173,645 & 3.66\\
   f & Death of Wang Yue & 13/Oct/2011 & 36,558 & 213,126 & 5.83 \\
   g & Guo Meimei Event & 21/Jun/2011 & 94,212 & 734,759 & 7.80 \\
   h & Qian Yunhui Event & 25/Dec/2010 & 35,054 & 260,720 & 7.44\\
   \hline
   \end{tabular}
%  \begin{flushleft}
%  \end{flushleft}
 \end{table*}

 The spreading channel also plays an important role in information spreading.  Generally, there are two ways for an individual to access information: (i) peer-to-peer communications via a social network; and (ii) an external influence from outside of the network.  Many previous studies traced the information spreading process by focusing on the interactions among individuals~\cite{Lu-Chen-Zhou-2011-NJP,Liben-Nowell-Kleinberg-2008-PNAS},
but spreading through the external channel was also found to be important~\cite{Goel-Watts-Goldstein-2012-EC,Proykova-Stauffer-2002-PA,Aral-Walker-2011-MS}.
In \emph{Twitter}, for example, about $71\%$ of information by volume can be attributed to internal diffusion within the network, and $29\%$ through external
influence~\cite{Myers-Zhu-Leskovec-2012-KDD}.  In innovation diffusion, Kocsis and Kun~\cite{Kocsis-Kun-2011-PRE} found a power-law with a crossover in the cluster size distribution, where the global effect due to the external channel determines the cluster's core and the local effect due to the internal channel
governs its growth.  There have also been studies on the effects of an external channel in epidemics, with transmission through a medium, e.g. mosquitoes, playing the role of an external channel, that an enhanced infection results from having multiple routes~\cite{Wang-Jin-Yang-Zhang-Zhou-Sun-2012-NA,Shi-Duan-Chen-2008-PA}. Although external influence can apparently enhance the information diffusion~\cite{Goel-Watts-Goldstein-2012-EC}, it remains unclear how the interplay between external influence and peer-to-peer  interactions affects information transmission in social networks~\cite{Myers-Zhu-Leskovec-2012-KDD}.

In this paper, we analyze internal and external influences on information spreading by tracking how events diffuse on the largest micro-blogging system -- \emph{Sina Weibo} (http://www.weibo.com/) -- in China.  Empirical results show that external influence plays a significant role, especially for events
that attract the media's attention readily at their immediate outbreaks. We then propose a diffusion model that incorporates both social interactions and media
effects~\cite{Goel-Watts-Goldstein-2012-EC} so as to illustrate the inter-relationship between the external and internal spreading channels. Both simulation and mathematical results of the model reveal that the spreading pattern is largely determined by the event's characteristic, as found in the empirical analyses.

\section{Empirical regularities}
\label{Sec:Results and Discussion}

As in other micro-blogging systems (e.g \emph{Twitter}), users of \emph{Sina Weibo} can post short messages, namely \emph{tweets}, in the variety of formats.  When an event occurs, there are basically two ways to learn about it.   Through the peer-to-peer interactions in a social network, referred to as internal influence, users receive automatically the contents posted by other users whom they follow.  Alternatively, users become aware of an event via an external influence outside the social network, e.g. via media broadcasts.

\begin{figure*}[htb]
  \centering
  \includegraphics[width=12cm]{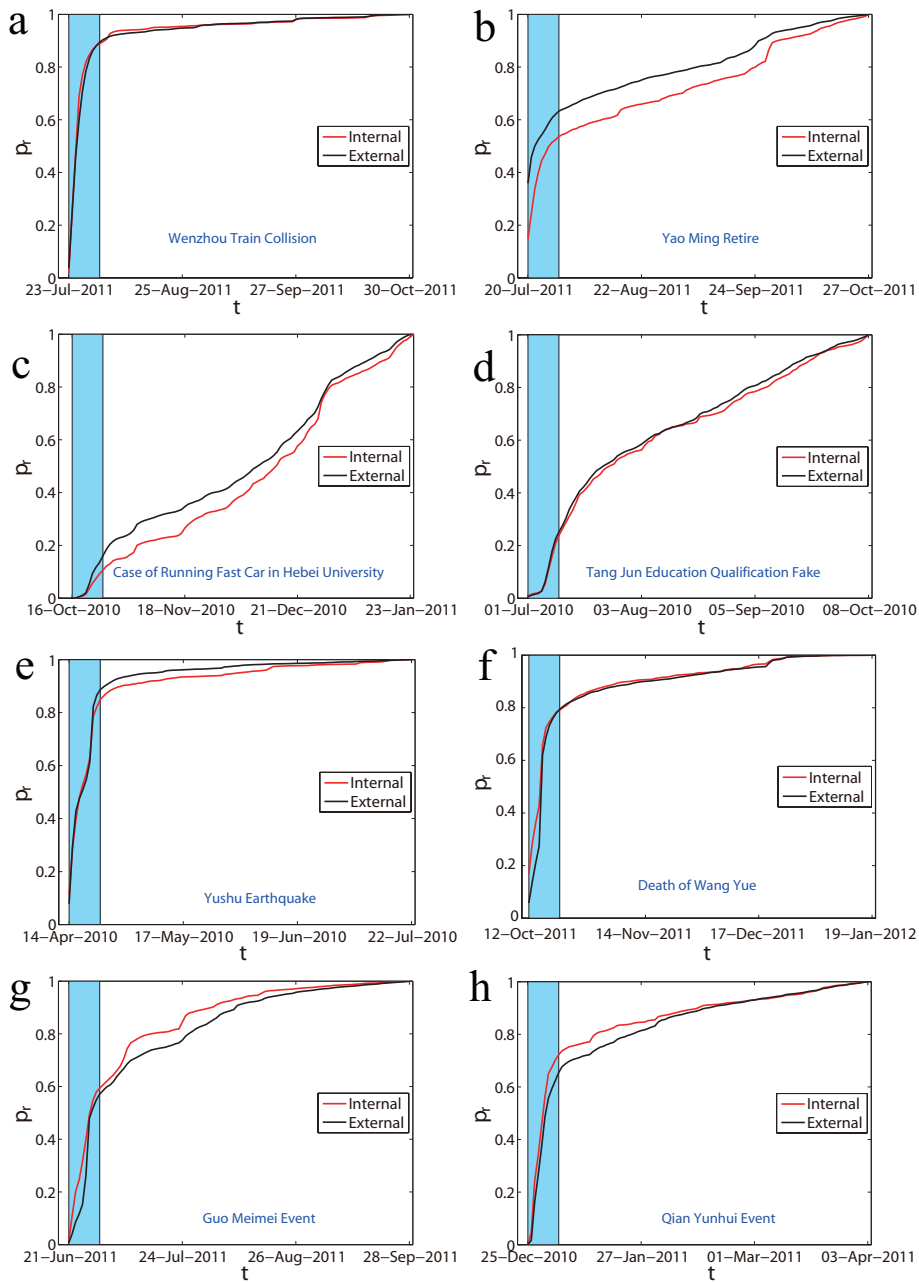}
  \caption{\label{Fig:Empirical:Dynamics} The spreading dynamics versus time of eight selected events on \emph{Sina Weibo}. Blue areas represent the spreading range within ten days after the corresponding events have occurred. Red and black curves represent the spreading affected by internal (retweets) and external influence (new tweets), respectively.
  }
\end{figure*}

Figure~\ref{Fig:Empirical:Dynamics} shows the spreading dynamics of some selected events from \emph{Sina Weibo} in the first 100 days of their outbreak. Details on the data are given in Supplementary Materials. Each topic carries at least $10^{4}$ new tweets or $10^{5}$ retweets, taken as a measure of the external and internal influences respectively. The basic statistic in Table~\ref{TB:Data} shows that the average retweet number is much larger than the new tweets, indicating that information diffusion on \emph{Sina Weibo} mainly through the internal channel, which is consistent with the results on \emph{Twitter} \cite{Myers-Zhu-Leskovec-2012-KDD}. Although all the events spread rapidly in the first ten days (shaped blue), the details of the spread patterns are different. In Fig. \ref{Fig:Empirical:Dynamics}, $p_r=\dfrac{n_{\#}(t)}{n_{\#}(t\rightarrow \infty)}$, where $n_{\#}(t)$ represents the cumulative number of messages posted through $\#$ (internal or external) channel till time $t$. Take the event labelled \textit{Yao Ming Retires} (Fig.~\ref{Fig:Empirical:Dynamics}b) for example. Being an internationally famous basketball star from China, people learned the news from media's coverage. The external influence led to a quicker outbreak of new tweets than retweets as the news propagated and was discussed ($p_r$ for external channel is higher than internal).  Another type of event can be observed in the example labelled the \textit{Guo Meimei Event} (Fig.~\ref{Fig:Empirical:Dynamics}g). It started when an ordinary lady showed off her wealthy lifestyle online and it did not draw the media's attention initially.  Many users gossiped when her account was revealed as a key official of the \emph{Chinese Red Cross}.  It became a hot topic quickly and eventually attracted the media's attention.  This strong internal influence led to a quicker outbreak of retweets as the item propagated and was discussed ($p_r$ for internal channel is higher). Figure~\ref{Fig:Empirical:Dynamics}b and Fig.~\ref{Fig:Empirical:Dynamics}g can be taken as typical of externally and internally initiated events, respectively, what are the events' characteristic mainly discussed in this work.

\begin{figure*}[htb]
  \centering
  \includegraphics[width=12cm]{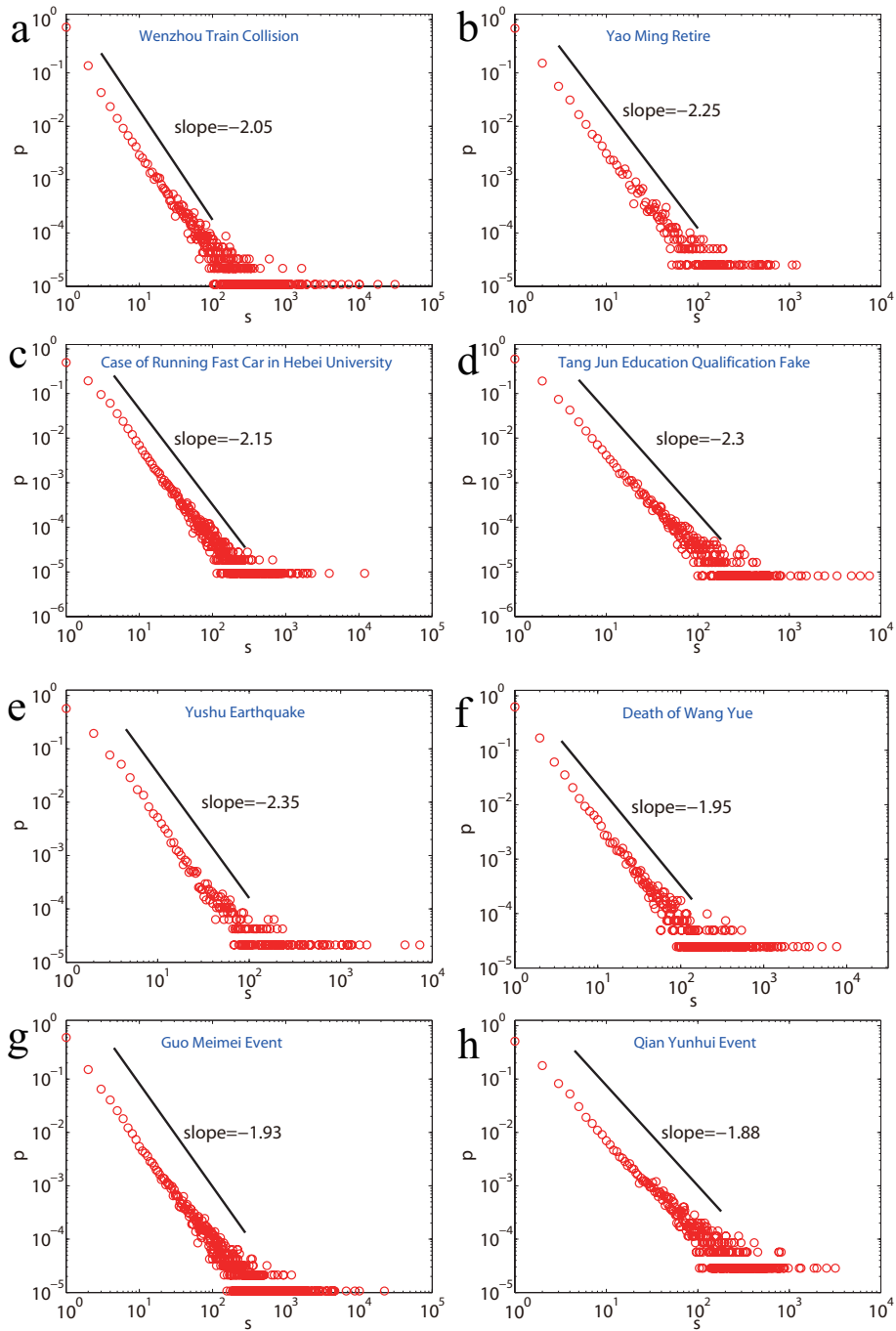}
  \caption{\label{Fig:Cascade:Size:Distribution:Empirical} The cascade size distribution for the diffusion of eight selected events. The distribution exponent is obtained by the Least Square Method.
  }
\end{figure*}

We further analyze the diffusion network of each tweet~\cite{Watts-2002-PNAS,Bakshy-Hofman-Mason-Watts-2011-WSDM}. It is a directed network with an edge $i$$\rightarrow$$j$ indicating information transmission from user $i$ to user $j$. A tweet can be traced from its origin through the \emph{retweeting
path} until the spreading terminates, showing the cascade due to the tweet.  The network consists entirely of internal channels and may be divided into serval unconnected communities due to effect of information blind areas~\cite{ZhangZK2014PO}.  For each event, the cascade size of each tweet can be found.
Figure~\ref{Fig:Cascade:Size:Distribution:Empirical} shows the spreading cascade size distribution for each event.  Each distribution exhibits a power-law with a slope around $-2.0$, similar to other systems~\cite{Goel-Watts-Goldstein-2012-EC}, and suggests the spreading dynamics via a few very large-scale cascade
and many small ones.  The details, however, are different for internally and externally initiated events.  For the \textit{Death of Wangyue} (Fig.~\ref{Fig:Cascade:Size:Distribution:Empirical}f), \textit{Guo Meimei} (Fig.~\ref{Fig:Cascade:Size:Distribution:Empirical}g) and \textit{Qian Yunhui} events (Fig.~\ref{Fig:Cascade:Size:Distribution:Empirical}h), the distribution exponents are less negative (smaller than 2), indicating events with stronger peer-to-peer interactions would lead to more larger-size cascades.  Furthermore, the average cascade size is also larger (see the metric $\langle N_r \rangle$ in Table~\ref{TB:Data}).  These events were initiated within the social network (see Fig.~\ref{Fig:Empirical:Dynamics}f-\ref{Fig:Empirical:Dynamics}h) until the media picked them up, and the discussions among peers gave rise to the large cascades.  In contrast, the other events caught the media's attention quickly.  The stronger external influence led to more message sources and smaller cascades (see Fig.~\ref{Fig:Cascade:Size:Distribution:Empirical}a-\ref{Fig:Cascade:Size:Distribution:Empirical}e), and thus a more negative exponent (larger than 2).

\section{Model Analyse}

\subsection{Model Description}

\begin{figure}[htb]
  \centering
  \includegraphics[width=8cm]{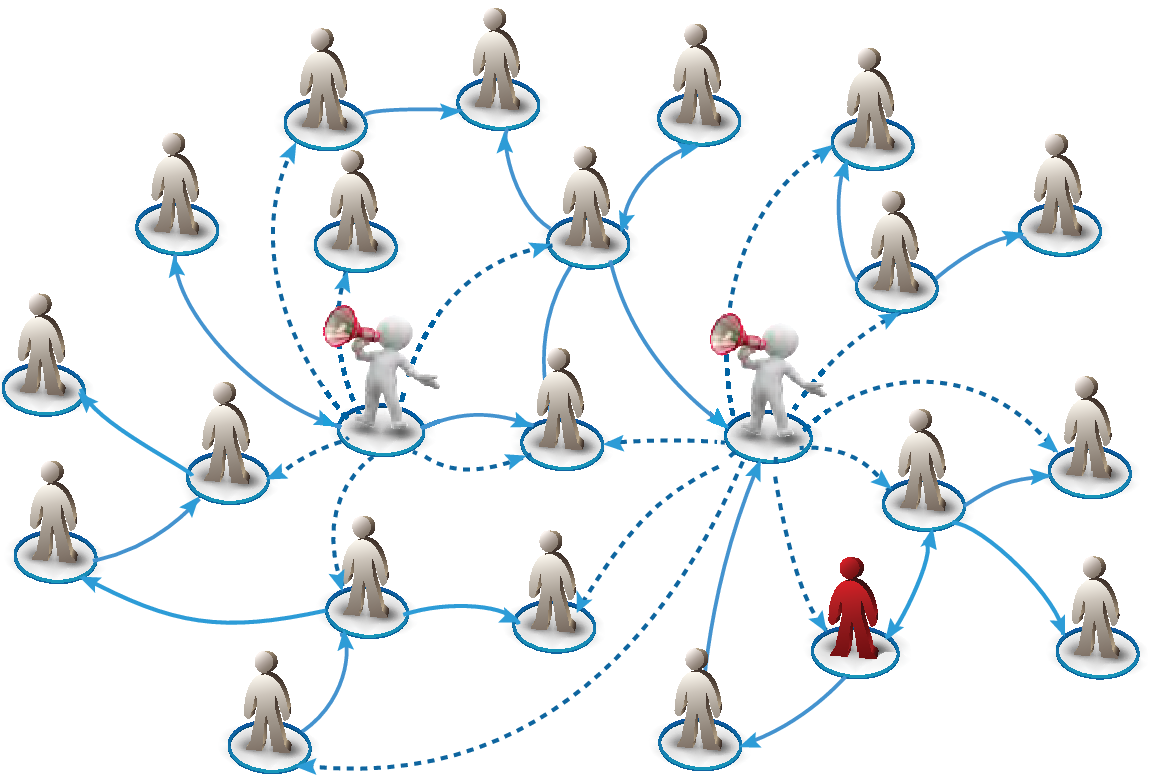}
  \caption{\label{Fig:Model} Illustration of information spreading model with both internal and external influence. The agents with loudspeakers represent the media-agents (external influence), which can spread information to the other agents with the same probability (dash arrows). Other gray agents represent ordinary individuals (the red agent is randomly selected to represent the information seed in the model), which can only deliver messages via peer-to-peer interactions based on existing social structure (solid arrows). All arrows indicate the direction of information flow.}
\end{figure}

We propose a theoretical model of information spreading that incorporates both internal and external influence.
Figure~\ref{Fig:Model} illustrates the model schematically.  Two types of agents -- ordinary individuals and media-agents -- are included in the network.  An agent receives information from another agent if s/he follows that agent, as indicated by the arrows (solid lines) for information flow.  A tiny fraction of
media-agents could broadcast information to the public represented by a group of agents (dashed lines) without them being followed in addition to forwarding information to followers.  We aim to incorporate (a) memory effects~\cite{Dodds-Watts-2004-PRL}; (b) external influences~\cite{Myers-Zhu-Leskovec-2012-KDD,Kocsis-Kun-2011-PRE}; and (c) non-redundancy of contacts~\cite{Lu-Chen-Zhou-2011-NJP}. As an event propagates, every agent takes on one of four states at any time: (a) \textit{unaware}: has not received information on event yet; (b) \textit{aware}: received information but hesitate to accept the content; (c) \textit{accepted}: accepted the content and ready to transmit it; (d) \textit{removed}: knew of the content but would not transmit it any more.  Therefore, an agent goes through the sequence of $unaware$$\rightarrow$$ aware $$ \rightarrow$$ accepted$$ \rightarrow$$ removed$, analogous to the SIR epidemic model.

The information diffusion process can be described as follows:

\begin{itemize}
\item To initiate an event, an agent is chosen randomly as a \textit{seed} (coloured red in Fig.~\ref{Fig:Model}) to spread the first piece of information, with the state set to \textit{accepted}.  All other agents are in the \textit{unaware} state.
\item At a time step $t$, every agent who turns into the \textit{accepted} state at the time step $(t-1)$ will post the information and become \textit{removed}.  For an ordinary agent, s/he forwards the information to her/his followers as a \textit{retweet}.  For a media-agent, the information is broadcasted as a \textit{new tweet} to a fraction of randomly chosen agents to mimic those who gather information from the media in addition to forwarding it as retweets to the followers.
\item At a time step $t$, all other agents check on information arrival.  For \textit{unaware} agents, they become \textit{aware} and evaluate a time-dependent \textit{acceptance probability} $p_{a}$ upon receipt of information according to the source (see Eq. (\ref{EQ:Exposure})). For \textit{aware} agents, they update $p_{a}$ if information arrives.  These agents then use $p_{a}$ to turn into \textit{accepted} at time $t$.  Those changed to the \textit{accepted} state are recorded.
\item  The steps are repeated until the information is spread to all accessible agents in the network.
\end{itemize}

There is a fraction ($0.1\%$ in this paper) of \textit{media-agents}, and each of them makes the same impact through broadcasting to $0.1\%$ of all agents.  The
\textit{acceptance probability} $p_{a}$ increases as one receives the same information repeatedly.  For an ordinary agent $i$ at time $t$, $p_a(i,t)$ is proportional to the amount of information $C(i,t)$ received so far and it is updated according to
\small
 \begin{equation}
 \label{EQ:Exposure}
 p_a(i,t) \propto C(i,t)=\left\{
 \begin{array}{lllll}
 C(i,t-1)+\sum\limits_{j\in \Gamma_{i_{t-1}}}w_{ji}~~~~~~~~~~~\mbox{if}~i\notin M_t\\\\
 C(i,t-1)+\sum\limits_{j\in \Gamma_{i_{t-1}}}w_{ji}+\beta ~~~\mbox{if}~i\in M_t
 \end{array},
 \right.
 \end{equation}
 \small
where $\Gamma_{i_{t-1}}$ is the set of agents that $i$ follows and who switches to the \textit{accepted} state at time $(t-1)$ and thus forward the information at time $t$ to $i$, $w_{ji}$ measures the internal influence due to interaction $j\rightarrow i$ ($w_{ji}=w$ is set for all pairs in the network), $\beta$ measures the external influence due to the media, and the set $M_t$ contains agents who received broadcasted information at time $t$.

For the acceptance probability $p_{a}^{(m)}$ of the \textit{media-agents}, we consider two extreme cases. For events initiated via gossips (labelled \textit{II} for internally initiated, such as the \textit{Guo Meimei} events) that the media are not eager to report, $p_{a}^{(m)} = p_{a}$ as in Eq.~(\ref{EQ:Exposure}) and thus follow the same updating rule. To mimic externally initiated (labelled \textit{EI}, such as the \textit{Yao Ming Retires}) events that the media rush to report, we set $p_{a}^{(m)} =1$ so that \textit{media-agents} accept the news immediately after they are \textit{aware} of the news. Note that Eq.~(\ref{EQ:Exposure}) incorporates the memory effect. Obviously, considering the external influence can enhance the information diffusion effect (see Supplementary Fig. S1).

\begin{figure}[htb]
  \centering
  \includegraphics[width=8cm]{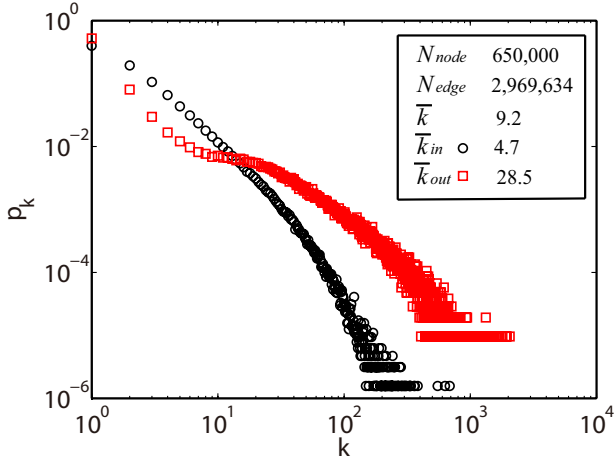}
  \caption{\label{Fig:Degree:Distribution} The degree distribution of the social network of \emph{Sina Weibo}. $k_{in}$ and $k_{out}$ represent the number of followers and followees for the corresponding user, respectively. The inset is the basic statistics of the original social network of \emph{Sina Weibo}. $N_{node}$ and $N_{edge}$ are the number of nodes and directed links, respectively. $\overline{k}$, $\overline{k}_{in}$ and $\overline{k}_{out}$ represent the average degree, average indegree and average outdegree, respectively. The nodes with zero indegree or outdegree are not counted. }
\end{figure}

\subsection{Simulation Results}

The model is implemented on the who-follow-whom online social network, i.e., \textit{followship network}, extracted from \textit{Sina Weibo} data.  The directed links give the direction of information flow, i.e., $i$$\rightarrow$$j$ when agent $i$ is followed by $j$. The basic statistics are given in Fig.~\ref{Fig:Degree:Distribution} (see inset).  The network reciprocity~\cite{Garlaschelli-Loffredo-2004-PRL} is about $15\%$. Fig. \ref{Fig:Degree:Distribution} shows the in-degree and out-degree distributions, excluding agents of degree zero. The distribution of $k_{out}$ is much broader than that of $k_{in}$, due to the two different social relationship in \emph{Sina Weibo}:
\emph{following} someone and \emph{being followed}.  Agents tend not to follow too many people due to their limited attention~\cite{Weng-Flammini-Vespignani-Menczer-2012-SR}. However, some targeted users, e.g. movie stars, are followed by a large number of agents without their consent.  The resulting mean degrees give $\overline{k}_{out} \gg \overline{k}_{in}$, suggesting that \textit{Sina Weibo} has developed into a structure
highly suitable for information flow (see Supplementary Fig. S2).

\begin{figure*}[htb]
  \centering
  \includegraphics[width=12cm]{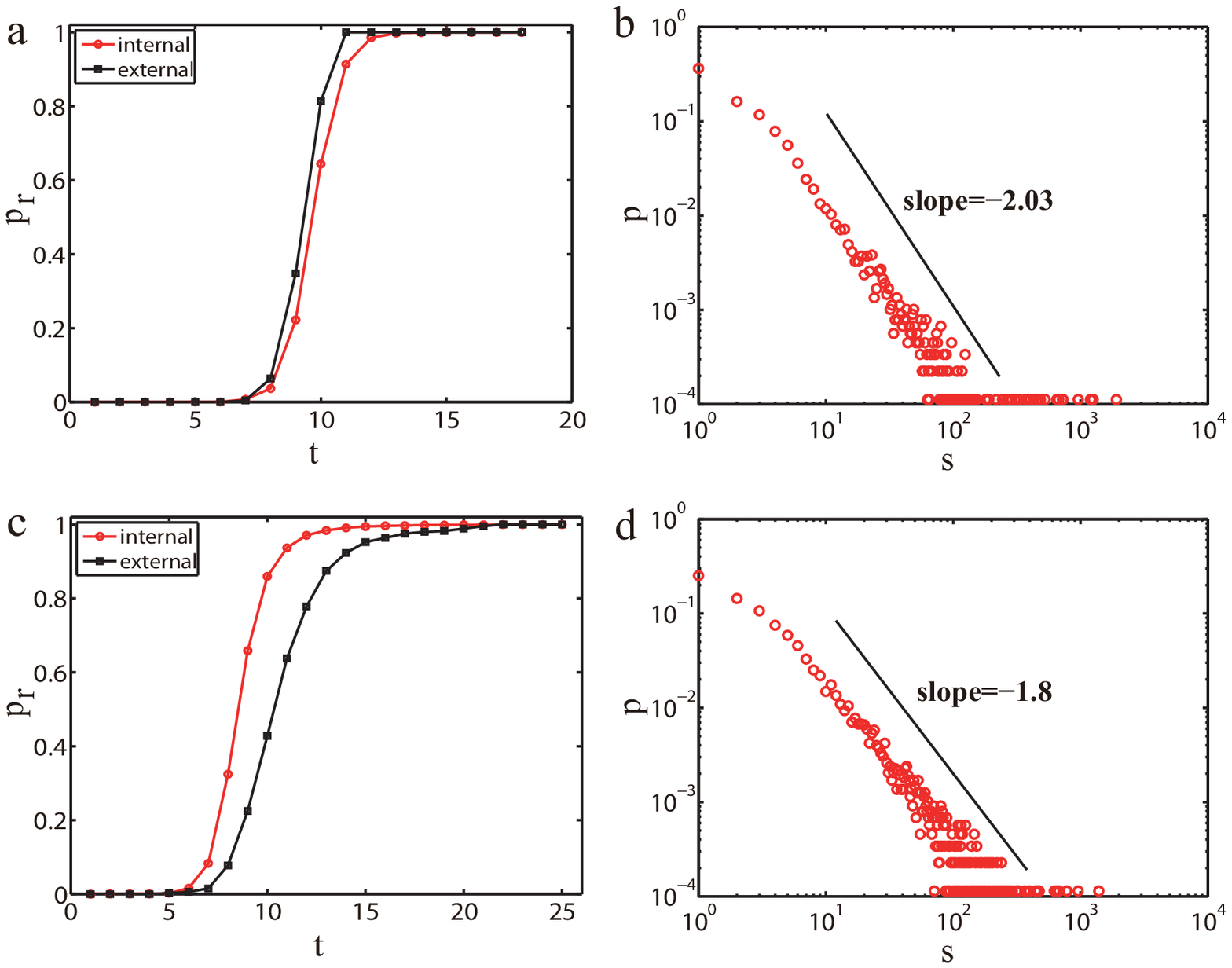}
  \caption{\label{Fig:Simulation} Simulation process of the information spreading via two different channels. {\bf{a}} and {\bf{c}}: cumulative fraction of removed individuals as a function of time; {\bf{b}} and {\bf{d}}: the cascade size distribution represented by the proposed model. The parameters are set as: {\bf{a}} and {\bf{b}}: $w=0.1$ and $\beta=0.01$ for the \textit{EI} events; {\bf{c}} and {\bf{d}}: $w=0.1$ and $\beta=0.01$ for the \textit{II} events.}
 \end{figure*}

\begin{figure*}[htb]
  \centering
  \includegraphics[width=12cm]{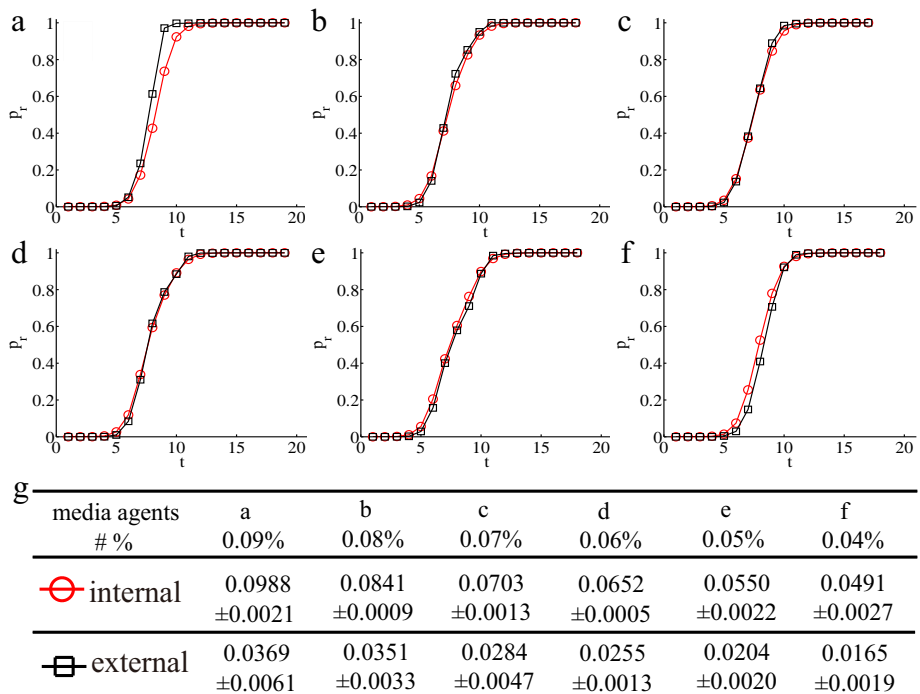}
  \caption{\label{Fig:Simulation:Mediaactors} Different patterns of information spreading via the two channels for various media-actor ratios for the \textit{EI} events. {\bf{a}}-{\bf{f}} represent the results for different ratios of media-agents ($\#\%$): {\bf{a}}~$0.09\%$, {\bf{b}}~$0.08\%$,~$\cdots$, {\bf{f}}~$0.04\%$ respectively. {\bf{g}} represents the fraction of removed individuals through different channels for various media-actor ratios. The average value and the corresponding standard deviation value are obtained by averaging over 100 independent realizations.}
\end{figure*}

\begin{figure*}[htb]
  \centering
  \includegraphics[width=12cm]{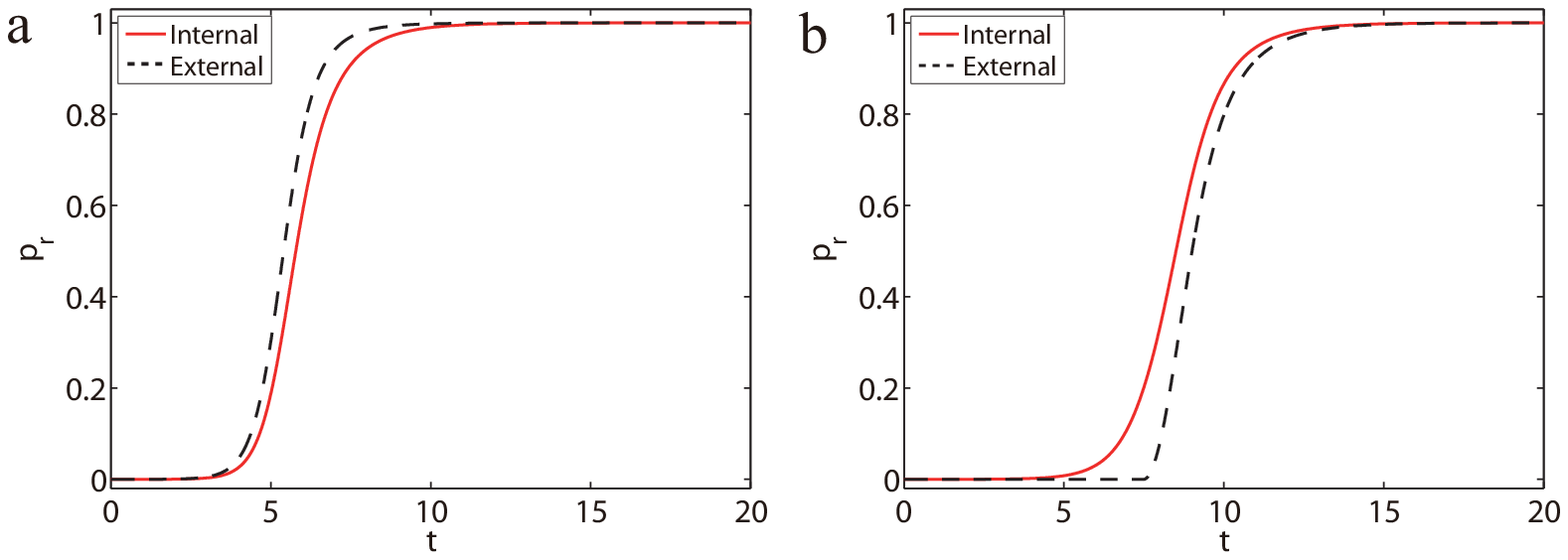}
  \caption{\label{Fig:Numerical:Analysis} Cumulative fraction of removed individuals versus time steps in the numerical analysis. {\bf{a}} for the \textit{EI} events; {\bf{b}} for the \textit{II} events. The parameters are set as: $w=0.1$ and $\beta=0.01$.
  }
\end{figure*}

We study both internally (\textit{II}) and externally initiated (\textit{EI}) events. As the empirical analysis, the fraction of followee-followers retweets and broadcasts (new tweets) are recorded as a function of time as the information spreads. Figure~\ref{Fig:Simulation} shows the results in terms of the cumulative fractions of removed agents due to the two processes for \textit{EI} (Fig.~\ref{Fig:Simulation}a) and \textit{II} events (Fig.~\ref{Fig:Simulation}c).  Tracing the propagation paths of many events, Fig.~\ref{Fig:Simulation}b and Fig. \ref{Fig:Simulation}d give the corresponding cascade size distributions.  Evidently, the model reproduces the key features in retweets and new tweets for \textit{EI} events (compare Fig. \ref{Fig:Simulation}a with Fig.~\ref{Fig:Empirical:Dynamics}a-\ref{Fig:Empirical:Dynamics}e and Fig. \ref{Fig:Simulation}b with Fig.~\ref{Fig:Cascade:Size:Distribution:Empirical}a-\ref{Fig:Cascade:Size:Distribution:Empirical}e), with $p_{r}(t)$ for new tweets higher than $p_{r}(t)$ for retweets and a more negative exponent in the cascade size distribution. Similarly, key features for $\textit{II}$ events are also reproduced (compare Fig. \ref{Fig:Simulation}c with Fig.~\ref{Fig:Empirical:Dynamics}f-\ref{Fig:Empirical:Dynamics}h and \ref{Fig:Simulation}d with Fig.~\ref{Fig:Cascade:Size:Distribution:Empirical}f-\ref{Fig:Cascade:Size:Distribution:Empirical}h), with $p_{r}(t)$ for retweets higher than $p_{r}(t)$ for new tweets and a less negative exponent in the cascade size distribution.

In order to further understand the effect of media-agents quantitatively, we detect the sensitivity of the proposed model to the ratio of media-agents. Figure ~\ref{Fig:Simulation:Mediaactors} shows the dynamics of the removed individuals through the two different channels for various media-agents ratios for the \textit{EI} events. Intriguingly, the spreading pattern can be apparently impacted by the ratio of media-agents, manifesting the burst attention changes from external channel for relatively large fraction of media-agents (Fig. \ref{Fig:Simulation:Mediaactors}a) to internal channel for small ones (Fig.~\ref{Fig:Simulation:Mediaactors}f). Thus, the external channel would only play the determining role in affecting the spreading patterns when there are enough media-agents in the systems for the \textit{EI} events (e.g. $0.06\%$ shown in Fig. \ref{Fig:Simulation:Mediaactors}d). In this way, only few media-agents would not be able to supersede the influence of gossips although they could response promptly to the \textit{EI} events. Therefore, it inspires that the information spreading patterns of the \textit{EI} events would be partially controlled by regulating the media-agents in real social networks, e.g. persuading ``\textit{stars}" not to forward the target message. However, different from \textit{EI} events, the information spread through the internal channel always bursts first for the \textit{II} events (see Supplementary Fig. S3). For such events, the media-agents can only influence information outbreak size, while unable to change the spreading patterns whatever how large they dominate the network.

\subsection{Mathematical Analysis}

In this section, we will give the mathematical analysis to illustrate the information diffusion patterns of the proposed model. We use superscript symbols $*^n$ and $*^m$ to represent the ordinary individuals and media-agents, respectively. Denote $S(t)$, $I(t)$ and $R(t)$ as the densities of \emph{aware- and unaware-}, \emph{accepted-} and \emph{removed-} states individuals. Adopting the mean-field approach
\cite{Pastor-Satorras-Vespignani-2001-PRL,Gomez-Gardenes-Latora-Moreno-Profumo-2008-PNAS,Pastor-Satorras-Castellano-Mieghem-Vespignani-2014-Arxiv},
we can obtain the differential equations describing the time evolution of the densities in each population:
\begin{small}
 \begin{equation}
 \setlength{\abovedisplayskip}{1mm}
 \setlength{\belowdisplayskip}{1mm}
 \label{EQ:S12}
 \left\{
 \begin{array}{l}
 \dfrac{dS^n(t)}{dt} =-p_a(t)S^n(t)I^n(t)\langle l \rangle -p_a(t)S^n(t)I^m(t)o\\[5pt]
 \dfrac{dI^n(t)}{dt} =p_a(t)S^n(t)I^n(t)\langle l \rangle +p_a(t)S^n(t)I^m(t)o-I^n(t)\\[5pt]
 \dfrac{dR^n(t)}{dt} =I^n(t)\\[5pt]
 \dfrac{dS^m(t)}{dt} =-p'_a(t)S^m(t)I^n(t)\langle l \rangle -p'_a(t)S^m(t)I^m(t)o\\[5pt]
 \dfrac{dI^m(t)}{dt} =p'_a(t)S^m(t)I^n(t)\langle l \rangle +p'_a(t)S^m(t)I^m(t)o-I^m(t)\\[5pt]
 \dfrac{dR^m(t)}{dt} =I^m(t)\\
 \end{array},
 \right.
 \end{equation}
 \end{small}
where $\langle l \rangle$ is the average out-degree of ordinary individuals, and $o$ is the number of agents that can receive the information through broadcasting of each media-agent, $p_a(t)$ and $p'_a(t)$ respectively are the \textit{accepted probability} for ordinary individuals and media-agents at time $t$.  According to Eq. (\ref{EQ:Exposure}), it can be obtained that the average $p_a$ is proportional to the number of \emph{removed State} individuals in
the system \cite{Valdez-Buono-Macri-Braunstein-2013-Fractal}. Therefore, we hereby assume the dynamics of $p_a(t)$ as the sigmoid function (also known as Fermi function in classic physics \cite{Traulsen-Pacheco-Nowak-2007-JTB}), $p_a(t) \sim \dfrac{c}{(1+e^{-at+b})}$ (Supplementary Fig. S1 and Fig. S2 show the plausibility to this hypothesis).

As we have illustrated in Model Description, for the diffusion of \textit{EI} events, the media-agents respond the event promptly, indicating that $p'_a=1$ all the time, while for the \textit{II} events, they are less attractive to media-agents when they happen, representing that $p'_a(t)$ for the media-agents are identical to ordinary individuals, saying $p'_a(t)=p_a(t)$. In addition, as there are only a small fraction of media-agents ($0.1\%$) are involved in the initial spreading process, resulting in $p'_a(t)\rightarrow 0$ in the initial times. Therefore, we can obtain the numerical results for Eq. (\ref{EQ:S12}) in Fig. \ref{Fig:Numerical:Analysis}, which share the similar pattern to the simulation and empirical results. That is to say, spreading via external channel is always ahead of that through internal channel for the diffusion of the \textit{EI} events (see Fig. \ref{Fig:Numerical:Analysis}a), and
vice verse for the \textit{II} events (see Fig. \ref{Fig:Numerical:Analysis}b). Further detailed analysis on the outbreak threshold of the proposed model is also presented in Supplementary Materials, and considering the external influence can diminish the information outbreak threshold significantly (see Supplementary Fig. S4).

\section{Conclusions \& Discussion}
\label{Sec:Conclusion}

In this paper, we have studied the internal and external influences on information transmission on social networks. Empirical analyses from a wide-range class of incidents of the Chinese largest social micro-blogging platform, \emph{Sina Weibo}, show that there are apparent differences between $EI$ and $II$ events. For the $EI$ events which attract more attention from media-agents would result in a broad and diverse popularity and corresponding large exponents of cascade size distribution. Comparatively, the $II$ events, mainly involved by social communications, show a very opposite phenomenon.  Therefore, the present findings demonstrate that the combination of out-of-network broadcasting and peer-to-peer interactions has played a significant role in facilitating the
emergence of different information transmission patterns.

In order to understand how information transmits with both peer-to-peer interactions and media effects, we have proposed an information spreading model based on the classical \emph{SIR} model, considering three representative characteristics: (i) memory effect; (ii) role of spreaders; and (iii) non-redundancy of contacts, which are all essential properties of the information diffusion and make it quite different from the basic models of biological epidemics. Thereinto, a small fraction of randomly selected individuals to act as the \textit{media-agents}, through which information can transmit out of the fixed structure of social network, referred to as the external influence. Both Simulation and mathematical results show that, though information diffusion depends largely  on the strength of the peer-to-peer interactions, the spreading pattern is essentially determined by the event attribute once the observed network structure is established, which agrees well with empirical analyses.

In the proposed model, individuals receive information via two approaches: internal (peer-to-peer contacts) and external (media) influences. The role of the external influence can be interpreted as two aspects: (i) the \textit{depth} effect: considered as the media's credibility, the amount of received
information of the \textit{aware}- and \textit{unaware}- state individuals, represented by the parameter $\beta$ in the model; (ii) the \textit{breadth} effect: considered as the media-agent's influence range, which brings more \textit{active} unaccepted individuals via media broadcasting. Besides, the internal and external influences would also promote the effects of each other. On one hand, the breadth effect of the external influence will arouse more active individuals to be aware of the information, and transmit the information to all their followers. On the other hand, events spread through the internal channel will attract more medias to report them, which additionally enlarges the external influence of the event diffusion. As a consequence, information will spread quicker and broader in social systems by the mutual reinforcement of external and internal influences (see Supplementary Fig. S1). Furthermore, we additionally observe the impact of network structure by investigating different media-agents ratios (see Fig. \ref{Fig:Simulation:Mediaactors} and Supplementary Fig. S3). It reveals that the population informed both from external and internal channels will increase with expanding the ratio of media-agents. In addition, the ratio of media-agents  would largely influence the spreading patterns for the \textit{EI} events. Therefore, strategy or policy makers should pay more attention to get along with the media-agents to obtain an effective way to manage the information diffusion.

The findings of this work may have various applications in studying how information spreads on social networks. (i) rumor spreading and detection are both very hot yet serious topics in purifying the air of public opinions; (ii) the field of information filtering confronts a huge challenge in dealing with tremendously increasing data every day, how to efficiently provide relevant information to users can be partially inspired to design more effective algorithms to obtain timely recommendations. The present work just provides a start point to preliminarily study the internal and external influences, a more comprehensive and in-depth understanding of multi-channel effects still need further efforts to discover.

\begin{acknowledgments}
We thank Prof. Jian-Hua Zhu for helpful comments and suggestions. This work was partially supported by Natural Science Foundation of China (Grant Nos. 11305043 and 11301490), Zhejiang Provincial Natural Science Foundation of China (Grant No. LY14A05001), Zhejiang Qianjiang Talents Project (QJC1302001), and the EU FP7 Grant 611272 (project GROWTHCOM).
\end{acknowledgments}

\bibliography{v21}

\begin{thebibliography}{48}%
\makeatletter
\providecommand \@ifxundefined [1]{%
 \@ifx{#1\undefined}
}%
\providecommand \@ifnum [1]{%
 \ifnum #1\expandafter \@firstoftwo
 \else \expandafter \@secondoftwo
 \fi
}%
\providecommand \@ifx [1]{%
 \ifx #1\expandafter \@firstoftwo
 \else \expandafter \@secondoftwo
 \fi
}%
\providecommand \natexlab [1]{#1}%
\providecommand \enquote  [1]{``#1''}%
\providecommand \bibnamefont  [1]{#1}%
\providecommand \bibfnamefont [1]{#1}%
\providecommand \citenamefont [1]{#1}%
\providecommand \href@noop [0]{\@secondoftwo}%
\providecommand \href [0]{\begingroup \@sanitize@url \@href}%
\providecommand \@href[1]{\@@startlink{#1}\@@href}%
\providecommand \@@href[1]{\endgroup#1\@@endlink}%
\providecommand \@sanitize@url [0]{\catcode `\\12\catcode `\$12\catcode
  `\&12\catcode `\#12\catcode `\^12\catcode `\_12\catcode `\%12\relax}%
\providecommand \@@startlink[1]{}%
\providecommand \@@endlink[0]{}%
\providecommand \url  [0]{\begingroup\@sanitize@url \@url }%
\providecommand \@url [1]{\endgroup\@href {#1}{\urlprefix }}%
\providecommand \urlprefix  [0]{URL }%
\providecommand \Eprint [0]{\href }%
\providecommand \doibase [0]{http://dx.doi.org/}%
\providecommand \selectlanguage [0]{\@gobble}%
\providecommand \bibinfo  [0]{\@secondoftwo}%
\providecommand \bibfield  [0]{\@secondoftwo}%
\providecommand \translation [1]{[#1]}%
\providecommand \BibitemOpen [0]{}%
\providecommand \bibitemStop [0]{}%
\providecommand \bibitemNoStop [0]{.\EOS\space}%
\providecommand \EOS [0]{\spacefactor3000\relax}%
\providecommand \BibitemShut  [1]{\csname bibitem#1\endcsname}%
\let\auto@bib@innerbib\@empty
%</preamble>
\bibitem [{\citenamefont {Chen}\ \emph
  {et~al.}(2013{\natexlab{a}})\citenamefont {Chen}, \citenamefont {Chen},
  \citenamefont {Gunnell},\ and\ \citenamefont
  {Yip}}]{Chen-Chen-Gunnell-Yip-2013-PlosOne}%
  \BibitemOpen
  \bibfield  {author} {\bibinfo {author} {\bibfnamefont {Y.-Y.}\ \bibnamefont
  {Chen}}, \bibinfo {author} {\bibfnamefont {F.}~\bibnamefont {Chen}}, \bibinfo
  {author} {\bibfnamefont {D.}~\bibnamefont {Gunnell}}, \ and\ \bibinfo
  {author} {\bibfnamefont {P.~S.~F.}\ \bibnamefont {Yip}},\ }\bibfield  {title}
  {\enquote {\bibinfo {title} {The impact of media reporting on the emergence
  of charcoal burning suicide in taiwan},}\ }\href@noop {} {\bibfield
  {journal} {\bibinfo  {journal} {PloS ONE}\ }\textbf {\bibinfo {volume} {8}},\
  \bibinfo {pages} {e55000} (\bibinfo {year} {2013}{\natexlab{a}})}\BibitemShut
  {NoStop}%
\bibitem [{\citenamefont {Zhang}\ \emph {et~al.}(2013)\citenamefont {Zhang},
  \citenamefont {Zhou}, \citenamefont {Zhang}, \citenamefont {Guan},\ and\
  \citenamefont {Zhou}}]{Zhang-Zhou-Zhang-Guan-Zhou-2013-PRE}%
  \BibitemOpen
  \bibfield  {author} {\bibinfo {author} {\bibfnamefont {Y.-C.}\ \bibnamefont
  {Zhang}}, \bibinfo {author} {\bibfnamefont {S.}~\bibnamefont {Zhou}},
  \bibinfo {author} {\bibfnamefont {Z.-Z.}\ \bibnamefont {Zhang}}, \bibinfo
  {author} {\bibfnamefont {J.-H.}\ \bibnamefont {Guan}}, \ and\ \bibinfo
  {author} {\bibfnamefont {S.-G.}\ \bibnamefont {Zhou}},\ }\bibfield  {title}
  {\enquote {\bibinfo {title} {Rumor evolution in social networks},}\
  }\href@noop {} {\bibfield  {journal} {\bibinfo  {journal} {Phys. Rev. E}\
  }\textbf {\bibinfo {volume} {87}},\ \bibinfo {pages} {032133} (\bibinfo
  {year} {2013})}\BibitemShut {NoStop}%
\bibitem [{\citenamefont {Moreno}\ \emph {et~al.}(2004)\citenamefont {Moreno},
  \citenamefont {Nekovee},\ and\ \citenamefont
  {Pacheco}}]{Moreno-Nekovee-Pacheco-2004-PRE}%
  \BibitemOpen
  \bibfield  {author} {\bibinfo {author} {\bibfnamefont {Y.}~\bibnamefont
  {Moreno}}, \bibinfo {author} {\bibfnamefont {M.}~\bibnamefont {Nekovee}}, \
  and\ \bibinfo {author} {\bibfnamefont {A.~F.}\ \bibnamefont {Pacheco}},\
  }\bibfield  {title} {\enquote {\bibinfo {title} {Dynamics of rumor spreading
  in complex networks},}\ }\href {\doibase 10.1103/PhysRevE.69.066130}
  {\bibfield  {journal} {\bibinfo  {journal} {Phys. Rev. E}\ }\textbf {\bibinfo
  {volume} {69}},\ \bibinfo {pages} {066130} (\bibinfo {year}
  {2004})}\BibitemShut {NoStop}%
\bibitem [{\citenamefont {Montanari}\ and\ \citenamefont
  {Saberi}(2010)}]{Montanari-Saberi-2010-PNAS}%
  \BibitemOpen
  \bibfield  {author} {\bibinfo {author} {\bibfnamefont {A.}~\bibnamefont
  {Montanari}}\ and\ \bibinfo {author} {\bibfnamefont {A.}~\bibnamefont
  {Saberi}},\ }\bibfield  {title} {\enquote {\bibinfo {title} {The spread of
  innovations in social networks},}\ }\href {\doibase 10.1073/pnas.1004098107}
  {\bibfield  {journal} {\bibinfo  {journal} {Proc. Natl. Acad. Sci. U.S.A.}\
  }\textbf {\bibinfo {volume} {107}},\ \bibinfo {pages} {20196--20201}
  (\bibinfo {year} {2010})}\BibitemShut {NoStop}%
\bibitem [{\citenamefont {Peres}(2014)}]{Peres2014PA}%
  \BibitemOpen
  \bibfield  {author} {\bibinfo {author} {\bibfnamefont {R.}~\bibnamefont
  {Peres}},\ }\bibfield  {title} {\enquote {\bibinfo {title} {The impact of
  networ characteristics on the diffusion of innovations},}\ }\href@noop {}
  {\bibfield  {journal} {\bibinfo  {journal} {Physica A}\ }\textbf {\bibinfo
  {volume} {402}},\ \bibinfo {pages} {330--343} (\bibinfo {year}
  {2014})}\BibitemShut {NoStop}%
\bibitem [{\citenamefont {Centola}(2010)}]{Centola-2010-Science}%
  \BibitemOpen
  \bibfield  {author} {\bibinfo {author} {\bibfnamefont {D.}~\bibnamefont
  {Centola}},\ }\bibfield  {title} {\enquote {\bibinfo {title} {The spread of
  behavior in an online social network experiment},}\ }\href {\doibase
  10.1126/science.1185231} {\bibfield  {journal} {\bibinfo  {journal}
  {Science}\ }\textbf {\bibinfo {volume} {329}},\ \bibinfo {pages} {1194--1197}
  (\bibinfo {year} {2010})}\BibitemShut {NoStop}%
\bibitem [{\citenamefont {Blansky}\ \emph {et~al.}(2013)\citenamefont
  {Blansky}, \citenamefont {Kavanaugh}, \citenamefont {Boothroyd},
  \citenamefont {Benson}, \citenamefont {Gallagher}, \citenamefont {Endress},\
  and\ \citenamefont
  {Sayama}}]{Blansky-Kavanaugh-Boothroyd-Benson-Gallagher-Endress-Sayama-2013-PLOSONE}%
  \BibitemOpen
  \bibfield  {author} {\bibinfo {author} {\bibfnamefont {D.}~\bibnamefont
  {Blansky}}, \bibinfo {author} {\bibfnamefont {C.}~\bibnamefont {Kavanaugh}},
  \bibinfo {author} {\bibfnamefont {C.}~\bibnamefont {Boothroyd}}, \bibinfo
  {author} {\bibfnamefont {B.}~\bibnamefont {Benson}}, \bibinfo {author}
  {\bibfnamefont {J.}~\bibnamefont {Gallagher}}, \bibinfo {author}
  {\bibfnamefont {J.}~\bibnamefont {Endress}}, \ and\ \bibinfo {author}
  {\bibfnamefont {H.}~\bibnamefont {Sayama}},\ }\bibfield  {title} {\enquote
  {\bibinfo {title} {Spread of academic success in a high school social
  network},}\ }\href@noop {} {\bibfield  {journal} {\bibinfo  {journal} {PLoS
  ONE}\ }\textbf {\bibinfo {volume} {8}},\ \bibinfo {pages} {e55944} (\bibinfo
  {year} {2013})}\BibitemShut {NoStop}%
\bibitem [{\citenamefont {Allen}\ \emph {et~al.}(2013)\citenamefont {Allen},
  \citenamefont {Weinrich}, \citenamefont {Hoppitt},\ and\ \citenamefont
  {Rendell}}]{Allen-Weinrich-Hoppitt-Rendell-2013-Science}%
  \BibitemOpen
  \bibfield  {author} {\bibinfo {author} {\bibfnamefont {J.}~\bibnamefont
  {Allen}}, \bibinfo {author} {\bibfnamefont {M.}~\bibnamefont {Weinrich}},
  \bibinfo {author} {\bibfnamefont {W.}~\bibnamefont {Hoppitt}}, \ and\
  \bibinfo {author} {\bibfnamefont {L.}~\bibnamefont {Rendell}},\ }\bibfield
  {title} {\enquote {\bibinfo {title} {Network-based diffusion analysis reveals
  cultural transmission of lobtail feeding in humpback whales},}\ }\href@noop
  {} {\bibfield  {journal} {\bibinfo  {journal} {Science}\ }\textbf {\bibinfo
  {volume} {340}},\ \bibinfo {pages} {485--488} (\bibinfo {year}
  {2013})}\BibitemShut {NoStop}%
\bibitem [{\citenamefont {Dybiec}\ \emph {et~al.}(2012)\citenamefont {Dybiec},
  \citenamefont {Mitarai},\ and\ \citenamefont
  {Sneppen}}]{Dybiec-Mitarai-Sneppen-2012-PRE}%
  \BibitemOpen
  \bibfield  {author} {\bibinfo {author} {\bibfnamefont {B.}~\bibnamefont
  {Dybiec}}, \bibinfo {author} {\bibfnamefont {N.}~\bibnamefont {Mitarai}}, \
  and\ \bibinfo {author} {\bibfnamefont {K.}~\bibnamefont {Sneppen}},\
  }\bibfield  {title} {\enquote {\bibinfo {title} {Information spreading and
  development of cultural centers},}\ }\href {\doibase
  10.1103/PhysRevE.85.056116} {\bibfield  {journal} {\bibinfo  {journal} {Phys.
  Rev. E}\ }\textbf {\bibinfo {volume} {85}},\ \bibinfo {pages} {056116}
  (\bibinfo {year} {2012})}\BibitemShut {NoStop}%
\bibitem [{\citenamefont {Nematzadeh}\ \emph {et~al.}(2014)\citenamefont
  {Nematzadeh}, \citenamefont {Ferrara}, \citenamefont {Flammini},\ and\
  \citenamefont {Ahn}}]{Nematzadeh-Ferrara-Flammini-Ahn-2014-PRL}%
  \BibitemOpen
  \bibfield  {author} {\bibinfo {author} {\bibfnamefont {A.}~\bibnamefont
  {Nematzadeh}}, \bibinfo {author} {\bibfnamefont {E.}~\bibnamefont {Ferrara}},
  \bibinfo {author} {\bibfnamefont {A.}~\bibnamefont {Flammini}}, \ and\
  \bibinfo {author} {\bibfnamefont {Y.-Y.}\ \bibnamefont {Ahn}},\ }\bibfield
  {title} {\enquote {\bibinfo {title} {Optimal network modularity for
  information diffusion},}\ }\href@noop {} {\bibfield  {journal} {\bibinfo
  {journal} {Phys. Rev. Lett.}\ }\textbf {\bibinfo {volume} {113}},\ \bibinfo
  {pages} {088701} (\bibinfo {year} {2014})}\BibitemShut {NoStop}%
\bibitem [{\citenamefont {Nagata}\ and\ \citenamefont
  {Shirayama}(2012)}]{Nagata-Shirayama-2012-PA}%
  \BibitemOpen
  \bibfield  {author} {\bibinfo {author} {\bibfnamefont {K.}~\bibnamefont
  {Nagata}}\ and\ \bibinfo {author} {\bibfnamefont {S.}~\bibnamefont
  {Shirayama}},\ }\bibfield  {title} {\enquote {\bibinfo {title} {Method of
  analyzing the influence of network structure on information diffusion},}\
  }\href {\doibase 10.1016/j.physa.2012.02.031} {\bibfield  {journal} {\bibinfo
   {journal} {Physica A}\ }\textbf {\bibinfo {volume} {391}},\ \bibinfo {pages}
  {3783--3791} (\bibinfo {year} {2012})}\BibitemShut {NoStop}%
\bibitem [{\citenamefont {Pastor-Satorras}\ and\ \citenamefont
  {Vespignani}(2001)}]{Pastor-Satorras-Vespignani-2001-PRL}%
  \BibitemOpen
  \bibfield  {author} {\bibinfo {author} {\bibfnamefont {R.}~\bibnamefont
  {Pastor-Satorras}}\ and\ \bibinfo {author} {\bibfnamefont {A.}~\bibnamefont
  {Vespignani}},\ }\bibfield  {title} {\enquote {\bibinfo {title} {Epidemic
  spreading in scale-free networks},}\ }\href {\doibase
  10.1103/PhysRevLett.86.3200} {\bibfield  {journal} {\bibinfo  {journal}
  {Phys. Rev. Lett.}\ }\textbf {\bibinfo {volume} {86}},\ \bibinfo {pages}
  {3200} (\bibinfo {year} {2001})}\BibitemShut {NoStop}%
\bibitem [{\citenamefont {Castellano}\ and\ \citenamefont
  {Pastor-Satorras}(2010)}]{Castellano-Pastor-Satorras-2010-PRL}%
  \BibitemOpen
  \bibfield  {author} {\bibinfo {author} {\bibfnamefont {C.}~\bibnamefont
  {Castellano}}\ and\ \bibinfo {author} {\bibfnamefont {R.}~\bibnamefont
  {Pastor-Satorras}},\ }\bibfield  {title} {\enquote {\bibinfo {title}
  {Thresholds for epidemic spreading in networks},}\ }\href {\doibase
  10.1103/PhysRevLett.105.218701} {\bibfield  {journal} {\bibinfo  {journal}
  {Phys. Rev. Lett.}\ }\textbf {\bibinfo {volume} {105}},\ \bibinfo {pages}
  {218701} (\bibinfo {year} {2010})}\BibitemShut {NoStop}%
\bibitem [{\citenamefont {Kitsak}\ \emph {et~al.}(2010)\citenamefont {Kitsak},
  \citenamefont {Gallos}, \citenamefont {Havlin}, \citenamefont {Liljeros},
  \citenamefont {Muchnik}, \citenamefont {Stanley},\ and\ \citenamefont
  {Makse}}]{Kitsak-Gallos-Havlin-Liljeros-Muchnik-Stanley-Makse-2010-NP}%
  \BibitemOpen
  \bibfield  {author} {\bibinfo {author} {\bibfnamefont {M.}~\bibnamefont
  {Kitsak}}, \bibinfo {author} {\bibfnamefont {L.}~\bibnamefont {Gallos}},
  \bibinfo {author} {\bibfnamefont {S.}~\bibnamefont {Havlin}}, \bibinfo
  {author} {\bibfnamefont {F.}~\bibnamefont {Liljeros}}, \bibinfo {author}
  {\bibfnamefont {L.}~\bibnamefont {Muchnik}}, \bibinfo {author} {\bibfnamefont
  {H.~E.}\ \bibnamefont {Stanley}}, \ and\ \bibinfo {author} {\bibfnamefont
  {H.}~\bibnamefont {Makse}},\ }\bibfield  {title} {\enquote {\bibinfo {title}
  {Identification of influential spreaders in complex networks},}\ }\href@noop
  {} {\bibfield  {journal} {\bibinfo  {journal} {Nat. Phys.}\ }\textbf
  {\bibinfo {volume} {11}},\ \bibinfo {pages} {888--893} (\bibinfo {year}
  {2010})}\BibitemShut {NoStop}%
\bibitem [{\citenamefont {Aral}(2011)}]{Aral-2011-MS}%
  \BibitemOpen
  \bibfield  {author} {\bibinfo {author} {\bibfnamefont {S.}~\bibnamefont
  {Aral}},\ }\bibfield  {title} {\enquote {\bibinfo {title} {Identifying social
  influence: a comment on opinion leadership and social contagion in new
  product diffusion},}\ }\href {\doibase 10.1016/j.physa.2007.11.048}
  {\bibfield  {journal} {\bibinfo  {journal} {Marketing Sci.}\ }\textbf
  {\bibinfo {volume} {30}},\ \bibinfo {pages} {217--223} (\bibinfo {year}
  {2011})}\BibitemShut {NoStop}%
\bibitem [{\citenamefont {Chen}\ \emph
  {et~al.}(2013{\natexlab{b}})\citenamefont {Chen}, \citenamefont {Gao},
  \citenamefont {Lu},\ and\ \citenamefont {Zhou}}]{Chen2013PO}%
  \BibitemOpen
  \bibfield  {author} {\bibinfo {author} {\bibfnamefont {D.-B.}\ \bibnamefont
  {Chen}}, \bibinfo {author} {\bibfnamefont {H.}~\bibnamefont {Gao}}, \bibinfo
  {author} {\bibfnamefont {L.-Y.}\ \bibnamefont {Lu}}, \ and\ \bibinfo {author}
  {\bibfnamefont {T.}~\bibnamefont {Zhou}},\ }\bibfield  {title} {\enquote
  {\bibinfo {title} {Identifying influential nodes in large-scale directed
  network: the role of clustering},}\ }\href@noop {} {\bibfield  {journal}
  {\bibinfo  {journal} {PLoS ONE}\ }\textbf {\bibinfo {volume} {8}},\ \bibinfo
  {pages} {e77455} (\bibinfo {year} {2013}{\natexlab{b}})}\BibitemShut
  {NoStop}%
\bibitem [{\citenamefont {Gross}\ \emph {et~al.}(2006)\citenamefont {Gross},
  \citenamefont {D'Lima},\ and\ \citenamefont
  {Blasius}}]{Gross-Dlima-Blasius-2006-PRL}%
  \BibitemOpen
  \bibfield  {author} {\bibinfo {author} {\bibfnamefont {T.}~\bibnamefont
  {Gross}}, \bibinfo {author} {\bibfnamefont {C.~J.~D.}\ \bibnamefont
  {D'Lima}}, \ and\ \bibinfo {author} {\bibfnamefont {B.}~\bibnamefont
  {Blasius}},\ }\bibfield  {title} {\enquote {\bibinfo {title} {Epidemic
  dynamics on an adaptive network},}\ }\href {\doibase
  10.1103/PhysRevLett.96.208701} {\bibfield  {journal} {\bibinfo  {journal}
  {Phys. Rev. Lett.}\ }\textbf {\bibinfo {volume} {96}},\ \bibinfo {pages}
  {208701} (\bibinfo {year} {2006})}\BibitemShut {NoStop}%
\bibitem [{\citenamefont {Zhou}\ and\ \citenamefont {Xia}(2014)}]{Zhou2014PA}%
  \BibitemOpen
  \bibfield  {author} {\bibinfo {author} {\bibfnamefont {Y.-Z.}\ \bibnamefont
  {Zhou}}\ and\ \bibinfo {author} {\bibfnamefont {Y.-J.}\ \bibnamefont {Xia}},\
  }\bibfield  {title} {\enquote {\bibinfo {title} {Epidemic spreading on
  weighted adaptive networks},}\ }\href@noop {} {\bibfield  {journal} {\bibinfo
   {journal} {Physica A}\ }\textbf {\bibinfo {volume} {399}},\ \bibinfo {pages}
  {16--23} (\bibinfo {year} {2014})}\BibitemShut {NoStop}%
\bibitem [{\citenamefont {Shen}\ \emph {et~al.}(2014)\citenamefont {Shen},
  \citenamefont {Wang}, \citenamefont {Fan}, \citenamefont {Di},\ and\
  \citenamefont {Lai}}]{Shen-Wang-Fan-Di-Lai-2014-NC}%
  \BibitemOpen
  \bibfield  {author} {\bibinfo {author} {\bibfnamefont {Z.-S.}\ \bibnamefont
  {Shen}}, \bibinfo {author} {\bibfnamefont {W.-X.}\ \bibnamefont {Wang}},
  \bibinfo {author} {\bibfnamefont {Y.}~\bibnamefont {Fan}}, \bibinfo {author}
  {\bibfnamefont {Z.-R.}\ \bibnamefont {Di}}, \ and\ \bibinfo {author}
  {\bibfnamefont {Y.-C.}\ \bibnamefont {Lai}},\ }\bibfield  {title} {\enquote
  {\bibinfo {title} {Reconstructing propagation networks with natural diversity
  and identifying hidden sources},}\ }\href@noop {} {\bibfield  {journal}
  {\bibinfo  {journal} {Nat. Commun.}\ }\textbf {\bibinfo {volume} {5}},\
  \bibinfo {pages} {4323} (\bibinfo {year} {2014})}\BibitemShut {NoStop}%
\bibitem [{\citenamefont {Gomez-Rodriguez}\ \emph {et~al.}(2013)\citenamefont
  {Gomez-Rodriguez}, \citenamefont {Leskovec},\ and\ \citenamefont
  {Sch{\"o}lkopf}}]{Gomez-Rodriguez-Leskovec-Scholkopf-2013-WSDM}%
  \BibitemOpen
  \bibfield  {author} {\bibinfo {author} {\bibfnamefont {M.}~\bibnamefont
  {Gomez-Rodriguez}}, \bibinfo {author} {\bibfnamefont {J.}~\bibnamefont
  {Leskovec}}, \ and\ \bibinfo {author} {\bibfnamefont {B.}~\bibnamefont
  {Sch{\"o}lkopf}},\ }\bibfield  {title} {\enquote {\bibinfo {title} {Structure
  and dynamics of information pathways in online media},}\ }in\ \href@noop {}
  {\emph {\bibinfo {booktitle} {Proc. 6th Int. Conf. WSDM}}}\ (\bibinfo
  {organization} {NewYork: ACM},\ \bibinfo {year} {2013})\ pp.\ \bibinfo
  {pages} {23--32}\BibitemShut {NoStop}%
\bibitem [{\citenamefont {Iribarren}\ and\ \citenamefont
  {Moro}(2012)}]{Iribarren-Moro-2009-PRL}%
  \BibitemOpen
  \bibfield  {author} {\bibinfo {author} {\bibfnamefont {J.~L.}\ \bibnamefont
  {Iribarren}}\ and\ \bibinfo {author} {\bibfnamefont {E.}~\bibnamefont
  {Moro}},\ }\bibfield  {title} {\enquote {\bibinfo {title} {Impact of human
  activity patterns on the dynamics of information diffusion},}\ }\href
  {\doibase 10.1103/PhysRevLett.103.038702} {\bibfield  {journal} {\bibinfo
  {journal} {Phys. Rev. Lett.}\ }\textbf {\bibinfo {volume} {103}},\ \bibinfo
  {pages} {038702} (\bibinfo {year} {2012})}\BibitemShut {NoStop}%
\bibitem [{\citenamefont {Karsai}\ \emph {et~al.}(2011)\citenamefont {Karsai},
  \citenamefont {Kivel{\"a}}, \citenamefont {Pan}, \citenamefont {Kaski},
  \citenamefont {Kert{\'e}sz}, \citenamefont {Barab{\'a}si},\ and\
  \citenamefont
  {Saram{\"a}ki}}]{Karsai-Kivela-Pan-Kaski-Kertesz-Barabasi-Saramaki-2011-PRE}%
  \BibitemOpen
  \bibfield  {author} {\bibinfo {author} {\bibfnamefont {M.}~\bibnamefont
  {Karsai}}, \bibinfo {author} {\bibfnamefont {M.}~\bibnamefont {Kivel{\"a}}},
  \bibinfo {author} {\bibfnamefont {R.~K.}\ \bibnamefont {Pan}}, \bibinfo
  {author} {\bibfnamefont {K.}~\bibnamefont {Kaski}}, \bibinfo {author}
  {\bibfnamefont {J.}~\bibnamefont {Kert{\'e}sz}}, \bibinfo {author}
  {\bibfnamefont {A.-L.}\ \bibnamefont {Barab{\'a}si}}, \ and\ \bibinfo
  {author} {\bibfnamefont {J.}~\bibnamefont {Saram{\"a}ki}},\ }\bibfield
  {title} {\enquote {\bibinfo {title} {Small but slow world: How network
  topology and burstiness slow down spreading},}\ }\href {\doibase
  10.1103/PhysRevE.83.025102} {\bibfield  {journal} {\bibinfo  {journal} {Phys.
  Rev. E}\ }\textbf {\bibinfo {volume} {84}},\ \bibinfo {pages} {046116(R)}
  (\bibinfo {year} {2011})}\BibitemShut {NoStop}%
\bibitem [{\citenamefont {Pinto}\ \emph {et~al.}(2012)\citenamefont {Pinto},
  \citenamefont {Thiran},\ and\ \citenamefont
  {Vetterli}}]{Pinto-Thiran-Vetterli-2012-PRL}%
  \BibitemOpen
  \bibfield  {author} {\bibinfo {author} {\bibfnamefont {P.~C.}\ \bibnamefont
  {Pinto}}, \bibinfo {author} {\bibfnamefont {P.}~\bibnamefont {Thiran}}, \
  and\ \bibinfo {author} {\bibfnamefont {M.}~\bibnamefont {Vetterli}},\
  }\bibfield  {title} {\enquote {\bibinfo {title} {Locating the source of
  diffusion in large-scale networks},}\ }\href {\doibase
  10.1103/PhysRevLett.109.068702} {\bibfield  {journal} {\bibinfo  {journal}
  {Phys. Rev. Lett.}\ }\textbf {\bibinfo {volume} {109}},\ \bibinfo {pages}
  {068702} (\bibinfo {year} {2012})}\BibitemShut {NoStop}%
\bibitem [{\citenamefont {Comin}\ and\ \citenamefont
  {Costa}(2011)}]{Comin-Costa-2011-PRE}%
  \BibitemOpen
  \bibfield  {author} {\bibinfo {author} {\bibfnamefont {C.~H.}\ \bibnamefont
  {Comin}}\ and\ \bibinfo {author} {\bibfnamefont {L.~F.}\ \bibnamefont
  {Costa}},\ }\bibfield  {title} {\enquote {\bibinfo {title} {Identifying the
  starting point of a spreading process in complex networks},}\ }\href
  {\doibase 10.1103/PhysRevE.84.056105} {\bibfield  {journal} {\bibinfo
  {journal} {Phys. Rev. E}\ }\textbf {\bibinfo {volume} {84}},\ \bibinfo
  {pages} {056105} (\bibinfo {year} {2011})}\BibitemShut {NoStop}%
\bibitem [{\citenamefont {Lloyd}\ and\ \citenamefont
  {May}(2001)}]{Lloyd-May-2001-Science}%
  \BibitemOpen
  \bibfield  {author} {\bibinfo {author} {\bibfnamefont {A.~L.}\ \bibnamefont
  {Lloyd}}\ and\ \bibinfo {author} {\bibfnamefont {R.~M.}\ \bibnamefont
  {May}},\ }\bibfield  {title} {\enquote {\bibinfo {title} {How viruses spread
  among computers and people},}\ }\href {\doibase 10.1209/0295-5075/82/38004}
  {\bibfield  {journal} {\bibinfo  {journal} {Science}\ }\textbf {\bibinfo
  {volume} {292}},\ \bibinfo {pages} {1316--1317} (\bibinfo {year}
  {2001})}\BibitemShut {NoStop}%
\bibitem [{\citenamefont {Newman}(2002)}]{Newman-2002-PRE}%
  \BibitemOpen
  \bibfield  {author} {\bibinfo {author} {\bibfnamefont {M.~E.~J.}\
  \bibnamefont {Newman}},\ }\bibfield  {title} {\enquote {\bibinfo {title}
  {Spread of epidemic disease on networks},}\ }\href {\doibase
  10.1103/PhysRevE.66.016128} {\bibfield  {journal} {\bibinfo  {journal} {Phys.
  Rev. E}\ }\textbf {\bibinfo {volume} {66}},\ \bibinfo {pages} {016128}
  (\bibinfo {year} {2002})}\BibitemShut {NoStop}%
\bibitem [{\citenamefont {Goel}\ \emph {et~al.}(2012)\citenamefont {Goel},
  \citenamefont {Watts},\ and\ \citenamefont
  {Goldstein}}]{Goel-Watts-Goldstein-2012-EC}%
  \BibitemOpen
  \bibfield  {author} {\bibinfo {author} {\bibfnamefont {S.}~\bibnamefont
  {Goel}}, \bibinfo {author} {\bibfnamefont {D.~J.}\ \bibnamefont {Watts}}, \
  and\ \bibinfo {author} {\bibfnamefont {D.~J.}\ \bibnamefont {Goldstein}},\
  }\bibfield  {title} {\enquote {\bibinfo {title} {The structure of online
  diffusion network},}\ }in\ \href@noop {} {\emph {\bibinfo {booktitle} {Proc.
  13th Int. Conf. EC}}}\ (\bibinfo {organization} {NewYork: ACM},\ \bibinfo
  {year} {2012})\ pp.\ \bibinfo {pages} {623--638}\BibitemShut {NoStop}%
\bibitem [{\citenamefont {L{\"u}}\ \emph {et~al.}(2011)\citenamefont {L{\"u}},
  \citenamefont {Chen},\ and\ \citenamefont {Zhou}}]{Lu-Chen-Zhou-2011-NJP}%
  \BibitemOpen
  \bibfield  {author} {\bibinfo {author} {\bibfnamefont {L.}~\bibnamefont
  {L{\"u}}}, \bibinfo {author} {\bibfnamefont {D.-B.}\ \bibnamefont {Chen}}, \
  and\ \bibinfo {author} {\bibfnamefont {T.}~\bibnamefont {Zhou}},\ }\bibfield
  {title} {\enquote {\bibinfo {title} {The small world yields the most
  effective information spreading},}\ }\href {\doibase
  10.1088/1367-2630/13/12/123005} {\bibfield  {journal} {\bibinfo  {journal}
  {New J. Phys.}\ }\textbf {\bibinfo {volume} {13}},\ \bibinfo {pages} {123005}
  (\bibinfo {year} {2011})}\BibitemShut {NoStop}%
\bibitem [{\citenamefont {Dodds}\ and\ \citenamefont
  {Watts}(2004)}]{Dodds-Watts-2004-PRL}%
  \BibitemOpen
  \bibfield  {author} {\bibinfo {author} {\bibfnamefont {P.~S.}\ \bibnamefont
  {Dodds}}\ and\ \bibinfo {author} {\bibfnamefont {D.~J.}\ \bibnamefont
  {Watts}},\ }\bibfield  {title} {\enquote {\bibinfo {title} {Universal
  behavior in a generalized model of contagion},}\ }\href {\doibase
  10.1073/PhysRevLett.92.218701} {\bibfield  {journal} {\bibinfo  {journal}
  {Phys. Rev. Lett.}\ }\textbf {\bibinfo {volume} {92}},\ \bibinfo {pages}
  {218701} (\bibinfo {year} {2004})}\BibitemShut {NoStop}%
\bibitem [{\citenamefont {Huang}\ \emph {et~al.}(2014)\citenamefont {Huang},
  \citenamefont {Li}, \citenamefont {Wang}, \citenamefont {Hua-Wei},
  \citenamefont {Li},\ and\ \citenamefont {Xue-Qi}}]{HuangJM2014SR}%
  \BibitemOpen
  \bibfield  {author} {\bibinfo {author} {\bibfnamefont {J.}~\bibnamefont
  {Huang}}, \bibinfo {author} {\bibfnamefont {C.}~\bibnamefont {Li}}, \bibinfo
  {author} {\bibfnamefont {W.-Q.}\ \bibnamefont {Wang}}, \bibinfo {author}
  {\bibfnamefont {S.}~\bibnamefont {Hua-Wei}}, \bibinfo {author} {\bibfnamefont
  {G.}~\bibnamefont {Li}}, \ and\ \bibinfo {author} {\bibfnamefont
  {C.}~\bibnamefont {Xue-Qi}},\ }\bibfield  {title} {\enquote {\bibinfo {title}
  {Temporal scaling in information propagation},}\ }\href@noop {} {\bibfield
  {journal} {\bibinfo  {journal} {Sci. Rep.}\ }\textbf {\bibinfo {volume}
  {4}},\ \bibinfo {pages} {5334} (\bibinfo {year} {2014})}\BibitemShut
  {NoStop}%
\bibitem [{\citenamefont {Crane}\ and\ \citenamefont
  {Sornette}(2008)}]{Crane-Sornette-2008-PNAS}%
  \BibitemOpen
  \bibfield  {author} {\bibinfo {author} {\bibfnamefont {R.}~\bibnamefont
  {Crane}}\ and\ \bibinfo {author} {\bibfnamefont {D.}~\bibnamefont
  {Sornette}},\ }\bibfield  {title} {\enquote {\bibinfo {title} {Robust dynamic
  classes revealed by measuring the response function of social system},}\
  }\href {\doibase 10.1073/pnas.0803685105} {\bibfield  {journal} {\bibinfo
  {journal} {Proc. Natl. Acad. Sci. U.S.A.}\ }\textbf {\bibinfo {volume}
  {105}},\ \bibinfo {pages} {15649--15653} (\bibinfo {year}
  {2008})}\BibitemShut {NoStop}%
\bibitem [{\citenamefont {Wu}\ and\ \citenamefont
  {Huberman}(2007)}]{Wu-Huberman-2007-PNAS}%
  \BibitemOpen
  \bibfield  {author} {\bibinfo {author} {\bibfnamefont {F.}~\bibnamefont
  {Wu}}\ and\ \bibinfo {author} {\bibfnamefont {B.~A.}\ \bibnamefont
  {Huberman}},\ }\bibfield  {title} {\enquote {\bibinfo {title} {Novelty and
  collective attention},}\ }\href {\doibase 10.1073/pnas.0704916104} {\bibfield
   {journal} {\bibinfo  {journal} {Proc. Natl. Acad. Sci. U.S.A.}\ }\textbf
  {\bibinfo {volume} {104}},\ \bibinfo {pages} {17599--17601} (\bibinfo {year}
  {2007})}\BibitemShut {NoStop}%
\bibitem [{\citenamefont {Liben-Nowell}\ and\ \citenamefont
  {Kleinberg}(2008)}]{Liben-Nowell-Kleinberg-2008-PNAS}%
  \BibitemOpen
  \bibfield  {author} {\bibinfo {author} {\bibfnamefont {D.}~\bibnamefont
  {Liben-Nowell}}\ and\ \bibinfo {author} {\bibfnamefont {J.}~\bibnamefont
  {Kleinberg}},\ }\bibfield  {title} {\enquote {\bibinfo {title} {Tracing
  information flow on a global scale using internet chain-letter data},}\
  }\href@noop {} {\bibfield  {journal} {\bibinfo  {journal} {Proc. Natl. Acad.
  Sci. U.S.A.}\ }\textbf {\bibinfo {volume} {105}},\ \bibinfo {pages}
  {4633--4638} (\bibinfo {year} {2008})}\BibitemShut {NoStop}%
\bibitem [{\citenamefont {Proykova}\ and\ \citenamefont
  {Stauffer}(2002)}]{Proykova-Stauffer-2002-PA}%
  \BibitemOpen
  \bibfield  {author} {\bibinfo {author} {\bibfnamefont {A.}~\bibnamefont
  {Proykova}}\ and\ \bibinfo {author} {\bibfnamefont {D.}~\bibnamefont
  {Stauffer}},\ }\bibfield  {title} {\enquote {\bibinfo {title} {Social
  percolation and the influence of mass media},}\ }\href@noop {} {\bibfield
  {journal} {\bibinfo  {journal} {Physica A}\ }\textbf {\bibinfo {volume}
  {312}},\ \bibinfo {pages} {300--304} (\bibinfo {year} {2002})}\BibitemShut
  {NoStop}%
\bibitem [{\citenamefont {Aral}\ and\ \citenamefont
  {Walker}(2011)}]{Aral-Walker-2011-MS}%
  \BibitemOpen
  \bibfield  {author} {\bibinfo {author} {\bibfnamefont {S.}~\bibnamefont
  {Aral}}\ and\ \bibinfo {author} {\bibfnamefont {D.}~\bibnamefont {Walker}},\
  }\bibfield  {title} {\enquote {\bibinfo {title} {Creating social contagion
  through viral product design: a randomized trial of peer influence in
  networks},}\ }\href {\doibase 10.1287/mnsc.1110.1421} {\bibfield  {journal}
  {\bibinfo  {journal} {Manag. Sci.}\ }\textbf {\bibinfo {volume} {57}},\
  \bibinfo {pages} {1623--1639} (\bibinfo {year} {2011})}\BibitemShut {NoStop}%
\bibitem [{\citenamefont {Myers}\ \emph {et~al.}(2012)\citenamefont {Myers},
  \citenamefont {Zhu},\ and\ \citenamefont
  {Leskovec}}]{Myers-Zhu-Leskovec-2012-KDD}%
  \BibitemOpen
  \bibfield  {author} {\bibinfo {author} {\bibfnamefont {S.}~\bibnamefont
  {Myers}}, \bibinfo {author} {\bibfnamefont {C.-G.}\ \bibnamefont {Zhu}}, \
  and\ \bibinfo {author} {\bibfnamefont {J.}~\bibnamefont {Leskovec}},\
  }\bibfield  {title} {\enquote {\bibinfo {title} {Information diffusion and
  external influence in networks},}\ }in\ \href@noop {} {\emph {\bibinfo
  {booktitle} {Proc. 18th Int. conf. KDD}}}\ (\bibinfo {organization} {New
  York: ACM},\ \bibinfo {year} {2012})\ pp.\ \bibinfo {pages}
  {33--41}\BibitemShut {NoStop}%
\bibitem [{\citenamefont {Kocsis}\ and\ \citenamefont
  {Kun}(2011)}]{Kocsis-Kun-2011-PRE}%
  \BibitemOpen
  \bibfield  {author} {\bibinfo {author} {\bibfnamefont {G.}~\bibnamefont
  {Kocsis}}\ and\ \bibinfo {author} {\bibfnamefont {F.}~\bibnamefont {Kun}},\
  }\bibfield  {title} {\enquote {\bibinfo {title} {Competition of information
  channels in the spreading of innovations},}\ }\href {\doibase
  10.1103/PhysRevE.84.026111} {\bibfield  {journal} {\bibinfo  {journal} {Phys.
  Rev. E}\ }\textbf {\bibinfo {volume} {84}},\ \bibinfo {pages} {026111}
  (\bibinfo {year} {2011})}\BibitemShut {NoStop}%
\bibitem [{\citenamefont {Wang}\ \emph {et~al.}(2012)\citenamefont {Wang},
  \citenamefont {Jin}, \citenamefont {Yang}, \citenamefont {Zhang},
  \citenamefont {Zhou},\ and\ \citenamefont
  {Sun}}]{Wang-Jin-Yang-Zhang-Zhou-Sun-2012-NA}%
  \BibitemOpen
  \bibfield  {author} {\bibinfo {author} {\bibfnamefont {Y.}~\bibnamefont
  {Wang}}, \bibinfo {author} {\bibfnamefont {Z.}~\bibnamefont {Jin}}, \bibinfo
  {author} {\bibfnamefont {Z.-M.}\ \bibnamefont {Yang}}, \bibinfo {author}
  {\bibfnamefont {Z.-K.}\ \bibnamefont {Zhang}}, \bibinfo {author}
  {\bibfnamefont {T.}~\bibnamefont {Zhou}}, \ and\ \bibinfo {author}
  {\bibfnamefont {G.-Q.}\ \bibnamefont {Sun}},\ }\bibfield  {title} {\enquote
  {\bibinfo {title} {{Global analysis of an SIS model with an infective vector
  on complex networks}},}\ }\href {\doibase 10.1016/j.nonrwa.2011.07.033}
  {\bibfield  {journal} {\bibinfo  {journal} {Nonlinear Anal.}\ }\textbf
  {\bibinfo {volume} {13}},\ \bibinfo {pages} {543--557} (\bibinfo {year}
  {2012})}\BibitemShut {NoStop}%
\bibitem [{\citenamefont {Shi}\ \emph {et~al.}(2008)\citenamefont {Shi},
  \citenamefont {Duan},\ and\ \citenamefont {Chen}}]{Shi-Duan-Chen-2008-PA}%
  \BibitemOpen
  \bibfield  {author} {\bibinfo {author} {\bibfnamefont {H.-J.}\ \bibnamefont
  {Shi}}, \bibinfo {author} {\bibfnamefont {Z.-S.}\ \bibnamefont {Duan}}, \
  and\ \bibinfo {author} {\bibfnamefont {G.-R.}\ \bibnamefont {Chen}},\
  }\bibfield  {title} {\enquote {\bibinfo {title} {{An SIS model with infective
  medium on complex networks}},}\ }\href {\doibase 10.1016/j.physa.2007.11.048}
  {\bibfield  {journal} {\bibinfo  {journal} {Physica A}\ }\textbf {\bibinfo
  {volume} {387}},\ \bibinfo {pages} {2133--2144} (\bibinfo {year}
  {2008})}\BibitemShut {NoStop}%
\bibitem [{\citenamefont {Watts}(2002)}]{Watts-2002-PNAS}%
  \BibitemOpen
  \bibfield  {author} {\bibinfo {author} {\bibfnamefont {D.~J.}\ \bibnamefont
  {Watts}},\ }\bibfield  {title} {\enquote {\bibinfo {title} {A simple model of
  global cascades on random networks},}\ }\href {\doibase
  10.1073/pnas.082090499} {\bibfield  {journal} {\bibinfo  {journal} {Proc.
  Natl. Acad. Sci. U.S.A.}\ }\textbf {\bibinfo {volume} {99}},\ \bibinfo
  {pages} {5766--5771} (\bibinfo {year} {2002})}\BibitemShut {NoStop}%
\bibitem [{\citenamefont {Bakshy}\ \emph {et~al.}(2011)\citenamefont {Bakshy},
  \citenamefont {Hofman}, \citenamefont {Mason},\ and\ \citenamefont
  {Watts}}]{Bakshy-Hofman-Mason-Watts-2011-WSDM}%
  \BibitemOpen
  \bibfield  {author} {\bibinfo {author} {\bibfnamefont {S.}~\bibnamefont
  {Bakshy}}, \bibinfo {author} {\bibfnamefont {J.}~\bibnamefont {Hofman}},
  \bibinfo {author} {\bibfnamefont {W.~A.}\ \bibnamefont {Mason}}, \ and\
  \bibinfo {author} {\bibfnamefont {D.~J.}\ \bibnamefont {Watts}},\ }\bibfield
  {title} {\enquote {\bibinfo {title} {Everyone's an influencer: quantifying
  influence on twitter},}\ }in\ \href@noop {} {\emph {\bibinfo {booktitle}
  {Proc. 4th Int. Conf. WSDM}}}\ (\bibinfo {organization} {NewYork: ACM},\
  \bibinfo {year} {2011})\ pp.\ \bibinfo {pages} {65--74}\BibitemShut {NoStop}%
\bibitem [{\citenamefont {Zhang}\ \emph {et~al.}(2014)\citenamefont {Zhang},
  \citenamefont {Zhang}, \citenamefont {Han},\ and\ \citenamefont
  {Liu}}]{ZhangZK2014PO}%
  \BibitemOpen
  \bibfield  {author} {\bibinfo {author} {\bibfnamefont {Z.-K.}\ \bibnamefont
  {Zhang}}, \bibinfo {author} {\bibfnamefont {C.-X.}\ \bibnamefont {Zhang}},
  \bibinfo {author} {\bibfnamefont {X.-P.}\ \bibnamefont {Han}}, \ and\
  \bibinfo {author} {\bibfnamefont {C.}~\bibnamefont {Liu}},\ }\bibfield
  {title} {\enquote {\bibinfo {title} {Emergence of blind areas in information
  spreading},}\ }\href@noop {} {\bibfield  {journal} {\bibinfo  {journal} {PLoS
  ONE}\ }\textbf {\bibinfo {volume} {9}},\ \bibinfo {pages} {e95785} (\bibinfo
  {year} {2014})}\BibitemShut {NoStop}%
\bibitem [{\citenamefont {Garlaschelli}\ and\ \citenamefont
  {Loffredo}(2004)}]{Garlaschelli-Loffredo-2004-PRL}%
  \BibitemOpen
  \bibfield  {author} {\bibinfo {author} {\bibfnamefont {D.}~\bibnamefont
  {Garlaschelli}}\ and\ \bibinfo {author} {\bibfnamefont {M.~I.}\ \bibnamefont
  {Loffredo}},\ }\bibfield  {title} {\enquote {\bibinfo {title}
  {{Fitness-dependent topological properties of the world trade web}},}\ }\href
  {\doibase 10.1103/PhysRevLett.93.188701} {\bibfield  {journal} {\bibinfo
  {journal} {Phys. Rev. Lett.}\ }\textbf {\bibinfo {volume} {93}},\ \bibinfo
  {pages} {188701} (\bibinfo {year} {2004})}\BibitemShut {NoStop}%
\bibitem [{\citenamefont {Weng}\ \emph {et~al.}(2012)\citenamefont {Weng},
  \citenamefont {Flammini}, \citenamefont {Vespignani},\ and\ \citenamefont
  {Menczer}}]{Weng-Flammini-Vespignani-Menczer-2012-SR}%
  \BibitemOpen
  \bibfield  {author} {\bibinfo {author} {\bibfnamefont {L.}~\bibnamefont
  {Weng}}, \bibinfo {author} {\bibfnamefont {A.}~\bibnamefont {Flammini}},
  \bibinfo {author} {\bibfnamefont {A.}~\bibnamefont {Vespignani}}, \ and\
  \bibinfo {author} {\bibfnamefont {F.}~\bibnamefont {Menczer}},\ }\bibfield
  {title} {\enquote {\bibinfo {title} {Competition among memes in a world with
  limited attention},}\ }\href@noop {} {\bibfield  {journal} {\bibinfo
  {journal} {Sci. Rep.}\ }\textbf {\bibinfo {volume} {2}},\ \bibinfo {pages}
  {335} (\bibinfo {year} {2012})}\BibitemShut {NoStop}%
\bibitem [{\citenamefont {G{\'o}mez-Garde{\~n}es}\ \emph
  {et~al.}(2008)\citenamefont {G{\'o}mez-Garde{\~n}es}, \citenamefont {Latora},
  \citenamefont {Moreno},\ and\ \citenamefont
  {Profumo}}]{Gomez-Gardenes-Latora-Moreno-Profumo-2008-PNAS}%
  \BibitemOpen
  \bibfield  {author} {\bibinfo {author} {\bibfnamefont {J.}~\bibnamefont
  {G{\'o}mez-Garde{\~n}es}}, \bibinfo {author} {\bibfnamefont {V.}~\bibnamefont
  {Latora}}, \bibinfo {author} {\bibfnamefont {Y.}~\bibnamefont {Moreno}}, \
  and\ \bibinfo {author} {\bibfnamefont {E.}~\bibnamefont {Profumo}},\
  }\bibfield  {title} {\enquote {\bibinfo {title} {Spreading of sexually
  transmitted diseases in heterosexual populations},}\ }\href@noop {}
  {\bibfield  {journal} {\bibinfo  {journal} {Proc. Natl. Acad. Sci. U.S.A.}\
  }\textbf {\bibinfo {volume} {105}},\ \bibinfo {pages} {1399--1404} (\bibinfo
  {year} {2008})}\BibitemShut {NoStop}%
\bibitem [{\citenamefont {Pastor-Satorras}\ \emph {et~al.}(2014)\citenamefont
  {Pastor-Satorras}, \citenamefont {Castellano}, \citenamefont {Van~Mieghem},\
  and\ \citenamefont
  {Vespignani}}]{Pastor-Satorras-Castellano-Mieghem-Vespignani-2014-Arxiv}%
  \BibitemOpen
  \bibfield  {author} {\bibinfo {author} {\bibfnamefont {R.}~\bibnamefont
  {Pastor-Satorras}}, \bibinfo {author} {\bibfnamefont {C.}~\bibnamefont
  {Castellano}}, \bibinfo {author} {\bibfnamefont {P.}~\bibnamefont
  {Van~Mieghem}}, \ and\ \bibinfo {author} {\bibfnamefont {A.}~\bibnamefont
  {Vespignani}},\ }\href@noop {} {\enquote {\bibinfo {title} {Epidemic
  processes in complex networks},}\ } (\bibinfo {year} {2014}),\ \bibinfo
  {note} {arXiv: 1408.2701}\BibitemShut {NoStop}%
\bibitem [{\citenamefont {Valdez}\ \emph {et~al.}(2013)\citenamefont {Valdez},
  \citenamefont {Buono}, \citenamefont {Macri},\ and\ \citenamefont
  {Braunstein}}]{Valdez-Buono-Macri-Braunstein-2013-Fractal}%
  \BibitemOpen
  \bibfield  {author} {\bibinfo {author} {\bibfnamefont {L.~D.}\ \bibnamefont
  {Valdez}}, \bibinfo {author} {\bibfnamefont {C.}~\bibnamefont {Buono}},
  \bibinfo {author} {\bibfnamefont {P.~A.}\ \bibnamefont {Macri}}, \ and\
  \bibinfo {author} {\bibfnamefont {L.~A.}\ \bibnamefont {Braunstein}},\
  }\bibfield  {title} {\enquote {\bibinfo {title} {Social distancing strategies
  against disease spreading},}\ }\href@noop {} {\bibfield  {journal} {\bibinfo
  {journal} {Fractals}\ }\textbf {\bibinfo {volume} {21}},\ \bibinfo {pages}
  {1350019} (\bibinfo {year} {2013})}\BibitemShut {NoStop}%
\bibitem [{\citenamefont {Traulsen}\ \emph {et~al.}(2007)\citenamefont
  {Traulsen}, \citenamefont {Pacheco},\ and\ \citenamefont
  {Nowak}}]{Traulsen-Pacheco-Nowak-2007-JTB}%
  \BibitemOpen
  \bibfield  {author} {\bibinfo {author} {\bibfnamefont {A.}~\bibnamefont
  {Traulsen}}, \bibinfo {author} {\bibfnamefont {J.~M.}\ \bibnamefont
  {Pacheco}}, \ and\ \bibinfo {author} {\bibfnamefont {M.~A.}\ \bibnamefont
  {Nowak}},\ }\bibfield  {title} {\enquote {\bibinfo {title} {Pairwise
  comparison and selection temperature in evolutionary game dynamics},}\
  }\href@noop {} {\bibfield  {journal} {\bibinfo  {journal} {J. Theor. Biol.}\
  }\textbf {\bibinfo {volume} {246}},\ \bibinfo {pages} {522--529} (\bibinfo
  {year} {2007})}\BibitemShut {NoStop}%
\end{thebibliography}%

%\documentclass{article}
%\pagestyle{empty}
%\setcounter{page}{6}
%\setlength\textwidth{135.0pt}
%\usepackage{amsmath}
%\begin{document}
%\begin{gather*}
%  \begin{matrix}  0 &  1 \\ 1 &  0 \end{matrix}  \quad
%  \begin{pmatrix} 0 & -i \\ i &  0 \end{pmatrix} \\
%  \begin{bmatrix} 0 & -1 \\ 1 &  0 \end{bmatrix} \quad
%  \begin{Bmatrix} 1 &  0 \\ 0 & -1 \end{Bmatrix} \\
%  \begin{vmatrix} a &  b \\ c &  d \end{vmatrix} \quad
%  \begin{Vmatrix} i &  0 \\ 0 & -i \end{Vmatrix}
%\end{gather*}
%\end{document}

\end{document}